\documentclass[manuscript]{aastex63}
\usepackage{shortvrb}
\usepackage{graphicx}                    % For eps figures, newer & more powerfull
\usepackage{amssymb}                     % useful mathematical symbols
\usepackage{color}                       % For color text: \color command
\usepackage{url}                         % For breaking URLs easily trough lines
\begin{document}

\title{Determining the Variations of Ca-K index and Features using a Century Long Equal Contrast Images from Kodaikanal Observatory}

\author{Jagdev Singh, Muthu Priyal, and B.Ravindra,}

\affil{Indian Institute of Astrophysics,Koramangala, Bengaluru-560034,
      \email{jsingh@iiap.res.in}  
             }
\begin{abstract}

In the earlier analysis of Ca-K spectroheliograms obtained at Kodaikanal Observatory, the ``Good'' images were used to investigate variations in the chromosphere. Still, the contrast of the images varied on a day-to-day basis. We developed a new methodology to generate images to form a uniform time series. We adjusted each image's contrast until the FWHM of the normalized intensity distribution attained a value between 0.10 and 0.11. This methodology of the ``Equal Contrast technique'' is expected to compensate for the change of emulsion, development, contrast of the images due to centering of Ca-K line on the exit slit and sky transparency. Besides, this procedure will correct variations in density-to-intensity conversion for different images. We find that the correlation between sunspot and Ca-K line data improves by a large amount. For example, correlation coefficient (CC) between monthly averaged sunspots and Ca-K plage areas for the equal contrast data improves to 0.9 compared to 0.75 for the ``Good'' data with unequal contrast. The CC for equal contrast images improves to $\sim$0.78 from $\sim$0.46 for the ``Okay'' data with unequal contrast. Even the CC between the plage area and the daily sunspot number is 0.85 for 100 years of data. This methodology also permits us to study the variations in Enhanced, Active, and Quiet networks with time along with good accuracy for about a century, for the first time. Further, this procedure can be used to combine data from different observatories to make a long time series.   
\end{abstract}

%\keywords{Sun: activity -- Sun: chromosphere -- Sun: faculae, plages}

\section{Introduction}

The long term periodicity or quasi-periodicities of the sun's large and small scale magnetic fields can be investigated using the long term spectroscopic and imaging data. The sun's images in the Ca-K line can be used as a proxy to study the variations in magnetic fields as there is a strong spatial correlation between Ca-K and magnetic features such as plages and networks \citep{1955ApJ...121..349B, 1959ApJ...130..193H, 1959ApJ...130..366L}. At the Mount Wilson observatory (MWO) and the Kodaikanal observatory (KO) spectroheliograms were obtained in Ca-K and H-alpha wavelengths daily for about 100 years during the 20th century. These data sets provide a time series of sun’s images to study the long term variations of the solar magnetic field. 

The earlier analysis of the digitized data of Ca-K images used density-to-intensity conversion using the theoretical equation or mean curve as step-wedge calibration was not recorded on all the images obtained during the long period \citep{1996GeoRL..23.2169F, 1998ApJ...496..998W, 2009ApJ...698.1000E, 2009SoPh..255..239T, 2010SoPh..264...31B, 2014SoPh..289..137P, 2017SoPh..292...85P, 2016ApJ...827...87C, 2019SoPh..294..145C}.

Several observatories have adapted their procedures to correct the images for the limb darkening and instrumental effects. After completing the task of making the quiet chromosphere of uniform nature, \citet{2019SoPh..294..131P} displayed each image along with the intensity distribution curve of the same on the computer screen. By visual inspection of these images and examining the corresponding intensity distribution curves, the images are classified as ``Good'' and ``Okay'', to generate two separate time series. Those two-time series analysis showed very different results \citep{2019SoPh..294..131P}. The ``Good'' time series with uniform images indicated smooth variations of Ca-K plage areas and Ca-K index with solar cycle phase. In contrast, the time series of remaining ``Okay'' images showed a large amount of scatter in the Ca-K parameters. Further, an inspection of the ``Good'' images indicated that there are variations in the contrast of images on a day-to-day basis as well as and on long term basis. The effect of contrast variations in Ca-K images still needs to be studied on the small and large scale activities on long-term. 

\citet{2007ApJ...657.1137L} have found that the Ca-K line intensity at the centre of the sun does not vary with the solar cycle phase. This result implies that quiet chromosphere does not show the solar cycle or long term variations. But \citet{2017SoPh..292...85P} found that the intensity of the quiet network varies with the solar cycle but with a very small amplitude. Therefore, it is reasonable to assume that the quiet chromosphere's intensity distribution does not vary with time. \citet{1998ApJ...496..998W} showed that the intensity contrast of active regions is beyond the intensity contrast of quiet chromosphere.  \citet{2019SoPh..294..131P} found that in the normalized intensity distribution of the image, intensity contrast greater than 1.30 represents plages and enhanced network (EN), intensity contrast between 1.20 to 1.30 represents active network (AN), and intensity contrast between 1.10 to 1.20 represents quiet network (QN). The remaining pixels, forming a large area of the sun, with intensity contrast lying between about 0.9 to 1.10 values, approximately follows the Gaussian distribution representing a quiet chromosphere. The extended tail of the intensity distribution of the image indicate the active area of the chromosphere. 

The Gaussian part of the intensity distribution representing the background chromosphere is expected to remain the same irrespective of the solar cycle phase \citep{2007ApJ...657.1137L}. The area under the extended tail varies significantly with time and is expected to show solar cycle variations. Therefore, all the images obtained on the day-to-day basis are expected to show similar intensity distribution for the quiet background chromosphere. This implies that the contrast of the images, thereby, full width at half maximum (FWHM) of the intensity distribution of all the images obtained during long periods should remain approximately the same. There might be a small change in the total area occupied by the quiet chromosphere over the solar cycle. Still, the FWHM of the intensity distribution is likely to remain the same with minor variations, if any. But FWHM of the intensity distribution was found to be related to the contrast of the images and much larger for high contrast images than for images with very low contrast \citep{2019SoPh..294..131P}. Therefore, by making the FWHM of the intensity distribution the same for all the images, it is possible to make the contrast of the long-time series data the same. 

It may be noted that images need to be corrected for the limb darkening and instrumental effects effectively before correcting for the contrast of the obtained images so that the FWHM of the intensity distribution becomes uniform within the specified limits.  In this paper, we describe the methodology adopted to make the contrast of all the images uniform and compare the results of the present analysis with those of earlier analysis obtained using data without correcting for the intensity contrast of images. The adapted methodology is somewhat similar approach used in studies of the Mt. Wilson archive of Ca-K observations \citep{2016A&A...585A..40P, 2020ApJ...897..181B}.

\section{Data Preparation and Analysis}
%\vspace*{1.0cm}

In our earlier paper \citep{2019SoPh..294..131P}, we have divided the data into two groups depending on the image quality and intensity distribution. Out of these two, one forming a time series of uniform images termed as ``Good'' and the other of remaining images termed as ``Okay'' series. Some bad quality images were discarded. Even though we have selected normal contrast data to make uniform time series termed as ``Good'' but still the contrast of the images varied daily due to sky conditions, development of photographic film, and variations in visually setting the centre of the Ca-K line on the exit slit of the spectrograph \citep{2014SoPh..289..137P}. The contrast of the images also varied due to the emulsion change on a long-term basis. In the earlier analysis, we found that FWHM of the intensity distribution varied systematically for the ``Good'' (hereafter called ``P-Good'') images. The average decrease in FWHM of the intensity distribution from 1907 to 1950 and then increase indicates that the contrast of the images varied with time on a long-time basis, probably due to change in emulsion and on a day-to-day basis due to sky conditions. Whereas in the ``Okay'' (hereafter called ``P-Okay'') series, the FWHM of the intensity distribution showed two branches due to the inclusion of very high and very low contrast images in this data set. These two series were analyzed separately. Now also, we have decided to analyse the two series of data separately to compare the earlier results with those of the new methodology being adopted. Sometimes we may refer ``P-Good'' and ``P-Okay'' images as ``P-image'' in general. 

\vspace*{1.0cm}

\begin{figure}[!h]
\centerline{\includegraphics[width=0.7\textwidth]{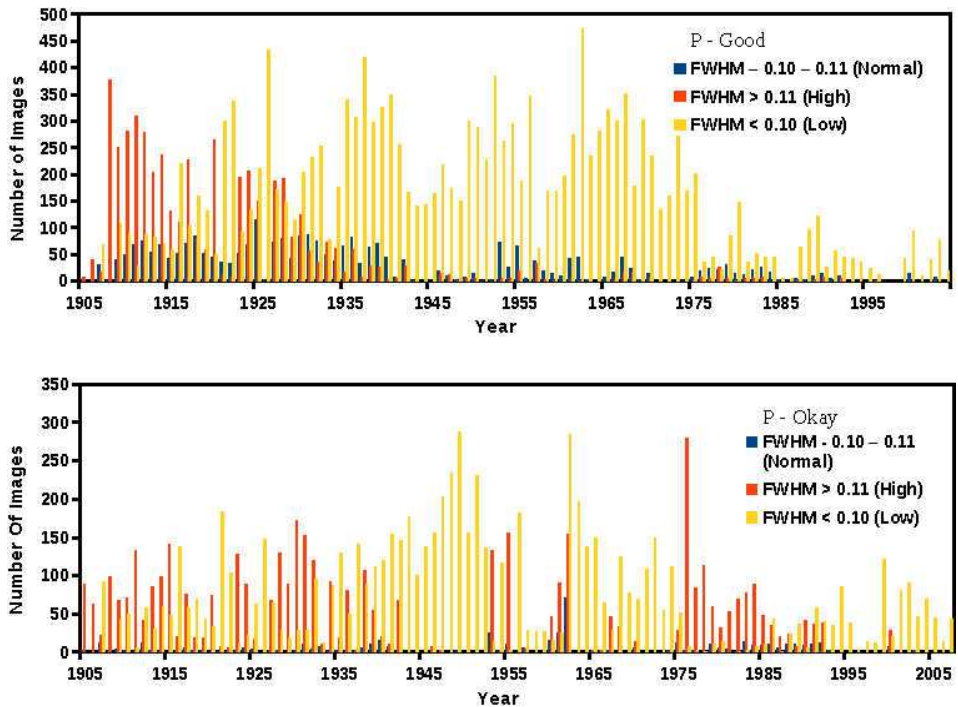}} 
\caption{The upper panel of the figure shows a histogram of the number of images per year with FWHM of the intensity distribution between 0.10 and 0.11, $>$ 0.11 and $<$ 0.1 in blue, red, and yellow lines, respectively for the images termed as ``P-Good'' images. The Lower panel indicates the same for ``P-Okay'' time series.}
\label{fig:1}
\end{figure}

We have determined the FWHM of the Gaussian fit to the intensity distribution of each image of the ``P-Good'' and ``P-Okay'' time series. The upper panel of Figure~\ref{fig:1} shows the histogram of a number of ``P-Good'' images per year whose FWHM of intensity distribution lies between 0.10 and 0.11 (shown in blue colour). The number of images per year with FWHM $>$ 0.11 and with FWHM $<$ 0.10 is also shown in the same plot in red and yellow color bars. The lower panel of the same figure shows the histogram of the ``P-Okay'' time series. The histogram of ``P-Good'' images indicates that the number of images with high contrast is more than low contrast images at the beginning of observations until 1930.  But after 1930, the number of images with low contrast is more than the high contrast images. The histogram for ``P-Okay'' images shows a similar trend in general. Besides, it indicates that the number of high contrast images are more during the period 1975 -- 1990. This comparison is relative. It may be noted that the number of images in the ``P-Okay'' data with FWHM in the range of 0.10 -- 0.11 is less compared to that for in the ``P-Good'' data.

In the present analysis, we studied the data of FWHM for a large number of images selected at a random spread over several years. We noted that images with FWHM in the range of 0.10 to 0.11 show chromospheric features very well. Therefore, it was decided to keep the FWHM of the intensity distribution between 0.10 -- 0.11 for all the images by changing their contrast. We compute the intensities of the modified image using the relation,

\begin{equation}
 I_{ECI} = I_{p}^{\gamma}
\end{equation}

Where, $I_{ECI}$ is the intensity of a pixel of new image (Equal Contrast Image), $I_{p}$ is the intensity of the pixel of the corrected image \citep{2019SoPh..294..131P}, and $\gamma$ is the contrast of the image.

First, we assume gamma = 1.0 and then compute the FWHM of the intensity distribution of the image. The images with FWHM of the intensity distribution less than 0.10 are low contrast images. The contrast of these images ($\gamma$) was increased in steps of 0.025 until the value of the FWHM lie between 0.10 and 0.11. The step of 0.025 was chosen after doing several experiments with different step sizes on many images. For the images with FWHM of the intensity distribution larger than 0.11 (high contrast images), the gamma value was decreased in steps 0.025 until the value of FWHM of the distribution becomes between 0.10 and 0.11. We term this methodology as ``equal contrast technique'' and can be used anywhere provided the images have been properly corrected for limb darkening and instrumental effects. The new methodology has been applied to both the ``P-Good'' and the ``P-Okay'' time series. After changing their contrast, the new images are called hereafter as ``ECI-Good'' (Equal Contrast Image) and ``ECI-Okay'' or ``ECI-image'' in general.

\subsection{Comparison of ``P-images'' and ``ECI-images'' during the Quiet Phase}

\begin{figure}[h]
\centerline{\includegraphics[width=0.8\textwidth]{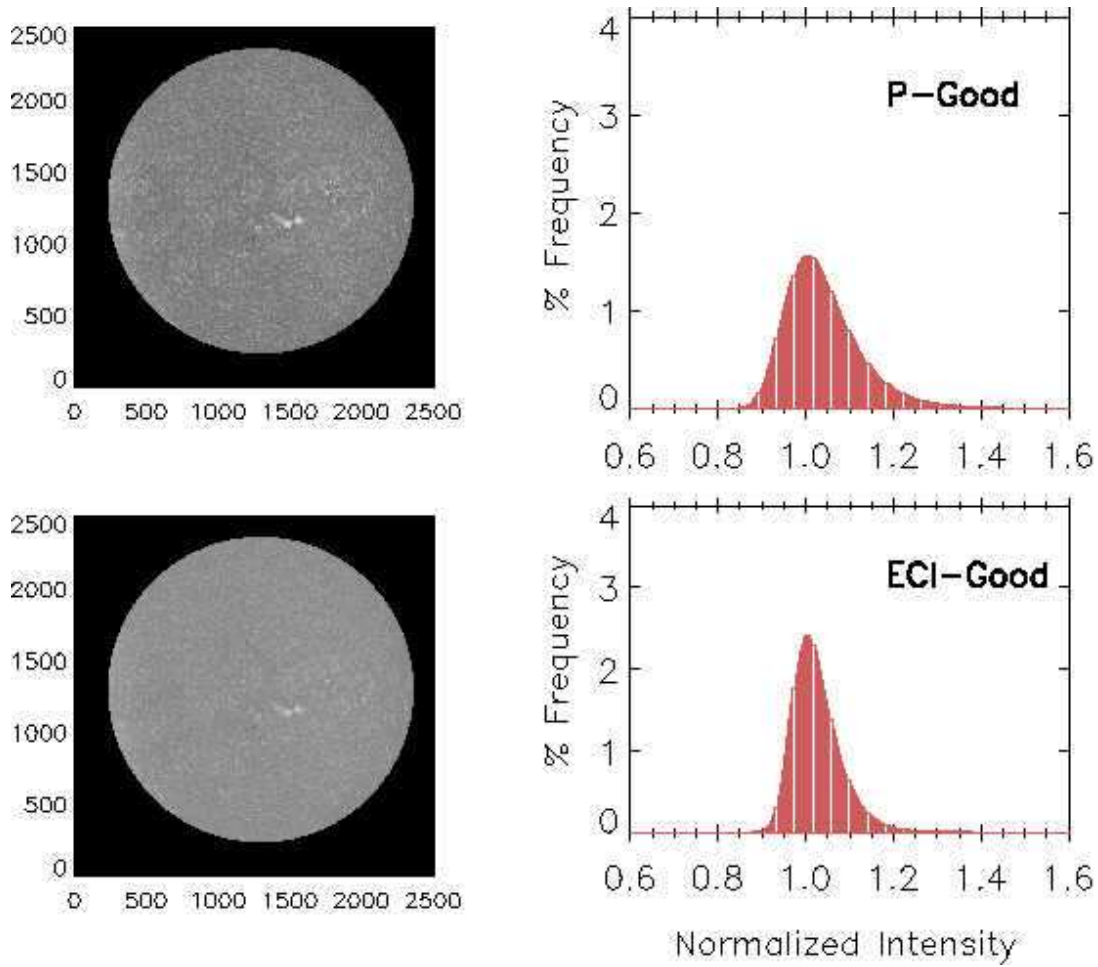}}
\caption{The analysed Ca-K image (``P-Good'') taken on 02 November 1911, during the quiet phase of the sun at Kodaikanal observatory indicates high contrast along with a broad intensity distribution curve in the top row. The bottom row shows the ``ECI-Good'' after decreasing the contrast such that the value of FWHM of the intensity distribution lies between 0.10 and 0.11.} 
\label{fig:2a}
\end{figure}

\begin{figure}[h]
\centerline{\includegraphics[width=0.8\textwidth]{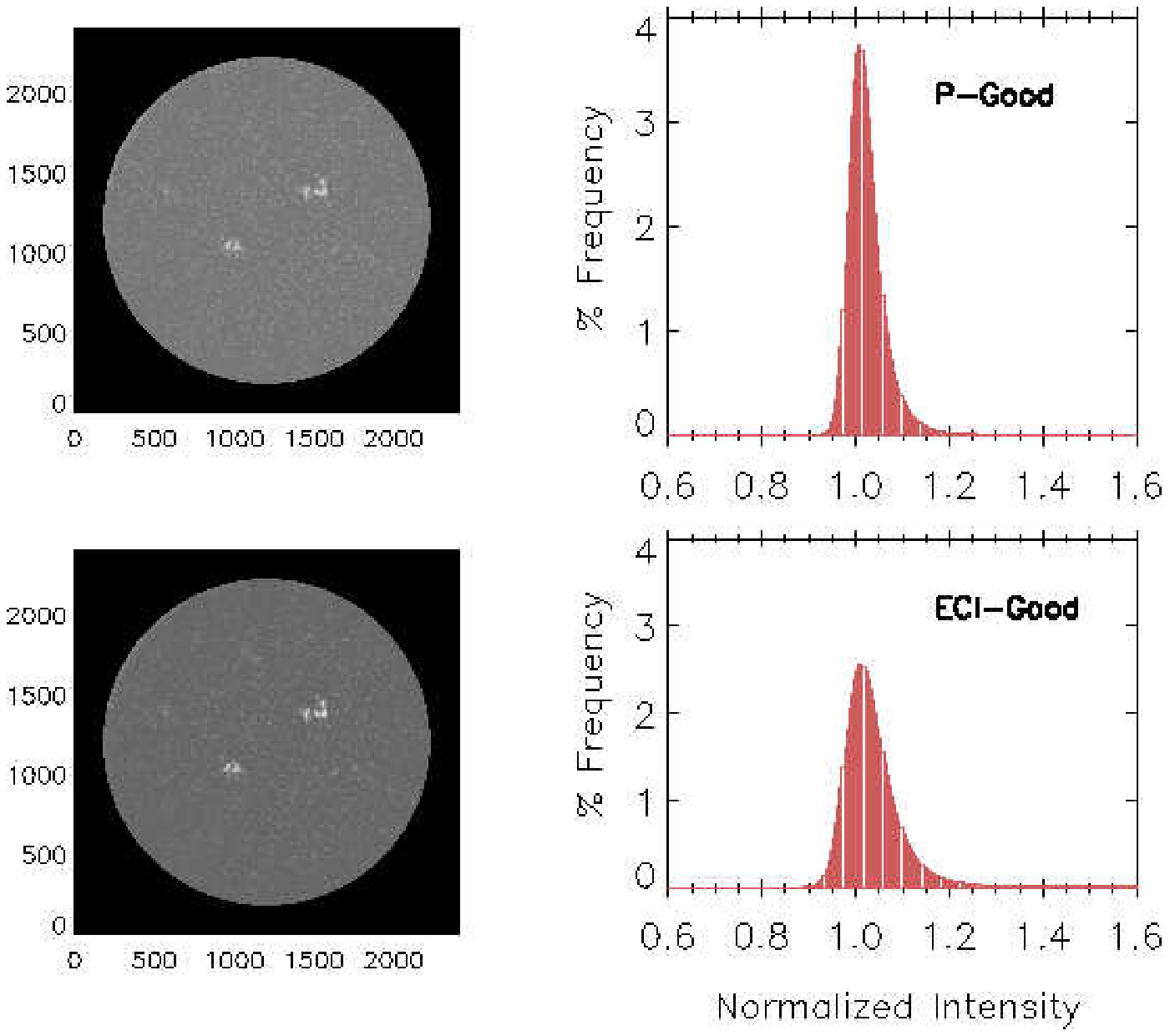}}
\caption{The analysed Ca-K ``P-Good'' image taken on 04 June 1923, during the quiet phase of the sun at Kodaikanal observatory shows very low contrast along with a narrow intensity distribution curve in the top row. The bottom row indicates the ``ECI-Good'' image and intensity distribution after increasing the contrast such that the value of FWHM of the intensity distribution lies between 0.10 and 0.11.}
\label{fig:2b}
\end{figure}

The left panel in the top row of Figure~\ref{fig:2a} shows a typical image belongs to the ``P-Good'' data set taken during the quiet phase of the solar cycle. The details of the image seen and FWHM of intensity distribution greater than 0.11 shown in the right-side panel imply high contrast. The FWHM of the intensity distribution is large, and the peak value of the frequency distribution is about 1.6\%. The extended tail beyond intensity contrast of 1.2 of the intensity distribution in the ``P-image'' indicates a significant existence of EN and AN even during the quiet phase of the sun, which is contrary to expectation. The left panel in the bottom row of this figure shows the ``ECI-image'' after adjusting the contrast of the image so that the FWHM of intensity distribution curve becomes between 0.10 and 0.11 (right-side panel). After changing the intensity contrast of the image, the tail of the intensity distribution curve reduces, indicating the less number of AN and almost absence of EN as expected during the quiet phase of the sun. Because of this procedure, the peak value of the frequency distribution increases to $\sim$2.5\%.

On the contrary, the top row of Figure~\ref{fig:2b} shows a low contrast image (also selected from the ``P-Good'' time series) taken during the quiet period of the solar cycle. This image exhibits a small FWHM and a high peak value ($\sim$3.8\%) in the frequency distribution. In the plot, due to the low contrast of the image, the intensity distribution $>$ 1.10 indicates an insignificant existence of QN. The bottom row of this figure shows the image after increasing the contrast such that FWHM of the intensity distribution lies between 0.10 -- 0.11. After increasing the contrast of the image, QN's contribution becomes significant approximately 2 -- 3 times. After adjusting the contrast of the images, both the high and low contrast (in ``P-Good'' time series) images acquire similar intensity distribution, as seen in the bottom row panels of Figures~\ref{fig:2a} and \ref{fig:2b}. A similar intensity distribution of the two types of images from the ``ECI-Good'' time series with a peak value of $\sim$2.5\% obtained during the quiet phase of the sun confirms the methodology adopted works well in both the cases. Hence, these two types of data can be used reliably for the study of short and long period variations in solar activity.

\subsection{Comparison of ``P-images'' and ``ECI-images'' during the Active Phase}

\begin{figure}[h]
\centerline{\includegraphics[width=0.8\textwidth]{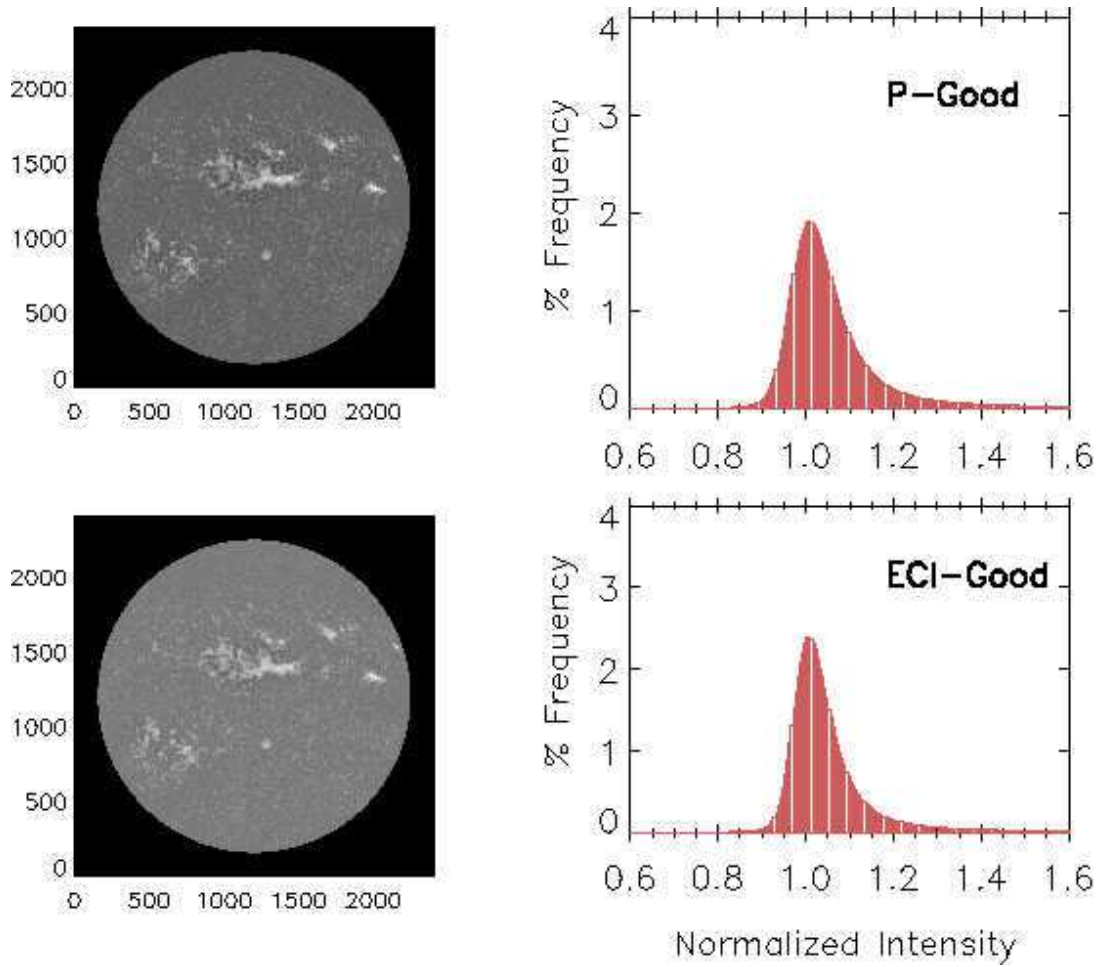}}
\caption{The analysed Ca-K ``P-Good'' image taken on 01 January 1929, during the active phase of the sun at Kodaikanal observatory along with the broad intensity distribution curve in the top row indicates very high contrast. The bottom row shows the ``ECI-Good'' of the same after decreasing the contrast such that the value of FWHM of the intensity distribution lies between 0.10 and 0.11.}
\label{fig:2c}
\end{figure}

\begin{figure}[h]
\centerline{\includegraphics[width=0.8\textwidth]{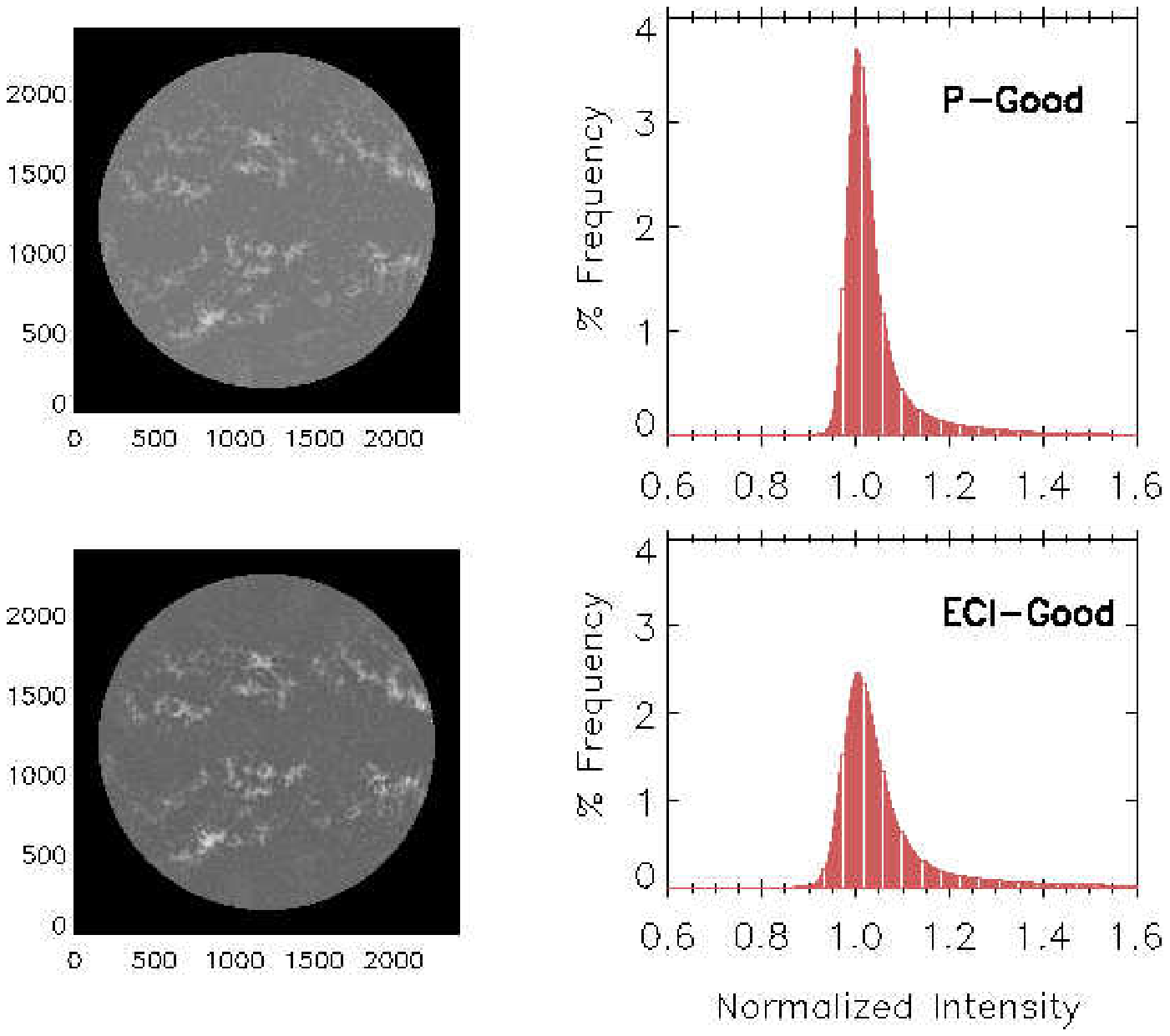}}
\caption{The analysed Ca-K line ``P-Good'' image taken on 01 March 1937, during the active phase of the sun at Kodaikanal Observatory, shows very low contrast along with a narrow intensity distribution curve in the top row. The bottom row indicates the ``ECI-Good'' after increasing the contrast such that the value of FWHM of the intensity distribution lies between 0.10 and 0.11.}
\label{fig:2d}
\end{figure}

Four panels of Figure~\ref{fig:2c} shows the images of P-Good (top) and ECI-Good (bottom) and their intensity distributions for a typical high contrast image obtained during the active phase of the solar cycle. The intensity contrast for part of the plage region exceeds~2, but we have plotted the intensity contrast up to 1.6 to clearly show intensity distribution. The intensity distribution of the image after adjusting the contrast appears realistic, indicating the area of the quiet chromosphere and the plage and EN regions’ existence. The peak value of the intensity distribution during the active phase appears to be less by a very small amount than that during the quiet phase to account for plage and EN regions’ existence. The area under the extended tail with an intensity contrast of more than 1.2 indicates the area occupied by the plages, EN, and AN networks. Four panels of Figure~\ref{fig:2d} shows the images of P-image and ECI-image after adjusting the contrast of the image and their intensity distributions representing low contrast image obtained during the active phase of the solar cycle. The tail of the distribution curve for ECI-image beyond 1.2 indicates the area of plages, EN, and AN.  

A Comparison of P-images in figures~\ref{fig:2a}, \ref{fig:2b}, \ref{fig:2c}, and \ref{fig:2d} indicates that FWHM of intensity distribution for the low contrast image is considerably less than that of high contrast and also the peak value of the frequency distribution is about 2 to 2.5 times more. In addition to this, the intensity distribution of the P-images of low contrast in Figures~\ref{fig:2b} and \ref{fig:2d} shows a significantly very less area under the extended tail representing plages, EN and AN. However, a large number of plages are visible in the P-image. After adjusting the contrast of the P-image, the area under the extended tail increases for ECI-image; thereby, the detected area of plages, EN, and AN increases. Similarly, after tuning the contrasts of high contrast P-images, intensity distributions of the ECI-images showed the area of detected plages, EN, and AN decreased. Thus, the values of detected plages, EN, and AN become realistic in both cases. A Comparison of all the intensity distributions of images after the fine-tune of contrast indicates that the peak values (~2.5\%) and FWHM become similar. This suggests that the area occupied by a quiet chromosphere may vary by a small percentage only in all the images. In contrast, the area under the extended tail varies significantly (percentage), showing the variations in active part of the sun.

The results of the analysis of the images in the ``P-Okay'' time series agrees well with the ``P-Good'' time series. Therefore, all the data can be combined and analysed together to reduce the gaps in the long time series to study solar variations with time.

\section{Results}
To study the chromosphere's long-term variations, we generally determine the Ca-K plage area, Ca-K index, and other such parameters with time. We also define the sum of plage, EN, AN and QN area as the ``total active area'' of the image. In other words, the ``total active area'' is the percentage area of an image occupied with intensity greater than 1.10 after normalizing the intensity of the image. It is also necessary to establish the correctness of the procedure adopted by comparing these derived parameters with the reliable other solar indices such as sunspots. First, we compare the recent measurements and similar data determined earlier with the sunspot data and then study the variations in other parameters of Ca-K line images such as networks.

\subsection{Comparison of daily Ca-K parameters with sunspot data for equal and non-equal contrast images}
\begin{figure}[h]
\centerline{\includegraphics[width=0.8\textwidth]{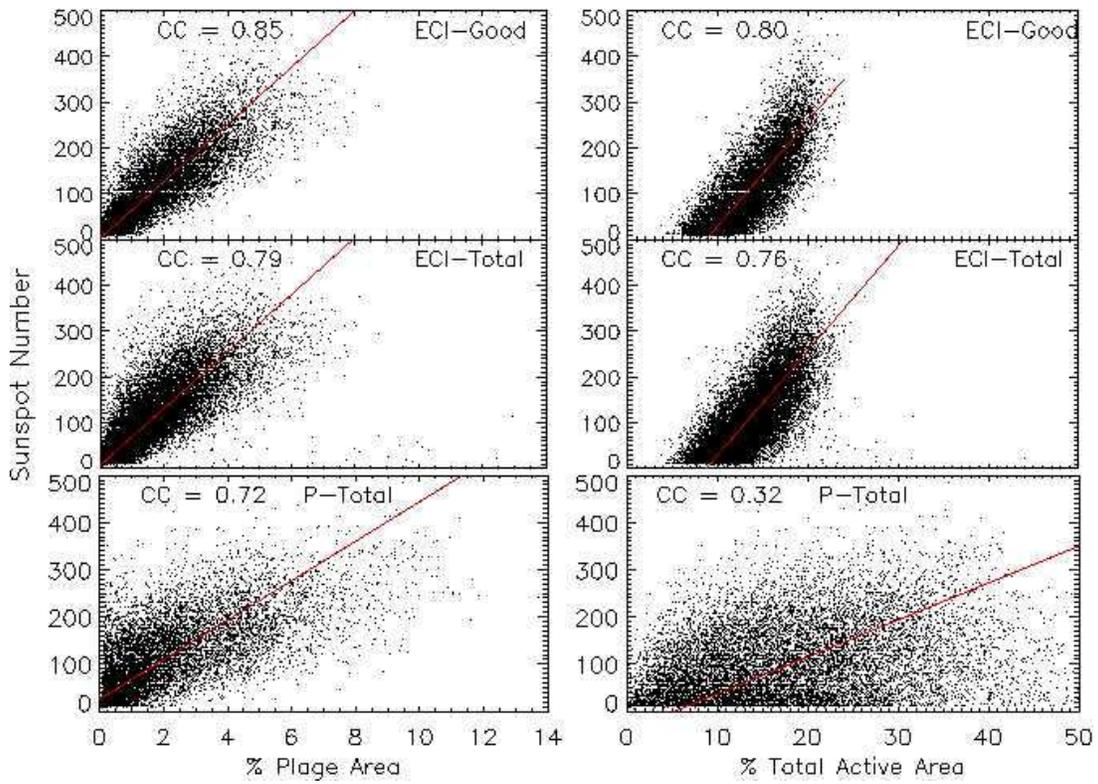}}
\caption{Two panels in the top-row show scatter plot of sunspot number versus plage area (left) and total active area (right) for the ``ECI-Good'' time series on daily basis. The middle row shows the same for the total data consisting of ``ECI-Good'' and ``ECI-Okay'' time series on daily basis. The bottom row shows the same for the total data of ``P-Good'' and ``P-Okay'' time series analysed by \citet{2019SoPh..294..131P}.}
\label{fig:3}
\end{figure}

To assess the improvement in the results due to equal contrast technique on the historical data, we first compare the results of this analysis, with some reliable works of earlier solar activity indices. Data on sunspot numbers and areas are available on a daily basis for the extended period for the comparison. The sunspot data obtained from various observatories were rescaled, combined them and formed a long daily time-series with negligible gap, if any. Thus, to compare the solar activity related indices, the sunspot data is most reliable. The left-side top panel of Figure~\ref{fig:3} shows a scatter plots between the SILSO Sunspot numbers (WDC-SILSO, Royal Observatory of Belgium, Brussels) versus identified plage areas using ECI-Good time series, whereas the right panel shows for the total active area on a daily basis. The middle two panels show the same for the total data i.e., combined data of the ECI-Good and ECI-Okay images, whereas two bottom panels show the same for combined P-images. 

\begin{table}[h]
\centering
%\begin{center}
\caption{Values of correlation coefficients between sunspots and Ca-K plage areas on daily basis. The ECI-Good, ECI-Combined and P-Combined represents the ``equal contrast images'' with good, good and okay (combined) and \citet{2019SoPh..294..131P} good and okay combined images.}
\begin{tabular}{|c| c| c|}
\hline
Data Type & Sunspot No. v/s plage areas & sunspot No. v/s total active area \\
ECI-Good & 0.85 & 0.8 \\
ECI-Combined & 0.79 & 0.76 \\
P-Combined & 0.72 & 0.32 \\
\hline
\end{tabular}
\label{tab:1}
\end{table}
%\end{center}

In Table~\ref{tab:1}, we list the values of correlation coefficients between sunspot data and Ca-K parameters derived by using the daily ECI images and earlier analysed daily images of \citet{2019SoPh..294..131P}. Table shows the values of correlation coefficients are large for the ECI-Good time series than others.
The correlation coefficient for the sunspot number and plage area of ECI-Good images is 0.85 and 0.80 for the total active area.  An excellent values of correlation coefficient for the data on a daily basis spread over about a 100-year period. The correlation coefficients for the combined ECI-images are marginally lower than those for ECI-Good data but still have a confidence level of greater than 99\%. The values of correlation coefficients for the combined P-images are significantly lower than those for the ECI-data. Also, there is a large scatter around the linear fit in scatter plots of P-images as compared to ECI-images. Further, in the case of the total active area, correlation coefficient of P-images is much less, 0.32 only as compared to 0.80 for the ECI-images. This indicates that the plage area can be determined to show the solar cycle variations with averages of the data over a long time, even with the non-uniform images of the time series. The maximum value of the total active region reaches up to 50\% in the case of P-images but decreases to about 25\% after tuning the contrast of images. The large value of the total active region may be due to high contrast images. This implies that it is almost impossible to study the periodic variations in the networks representing the small scale magnetic field with data of non-uniform time series.

\subsection{Comparison of monthly averaged Ca-K parameters with Sunspot Data}

\begin{figure}[h]
\centerline{\includegraphics[width=0.8\textwidth]{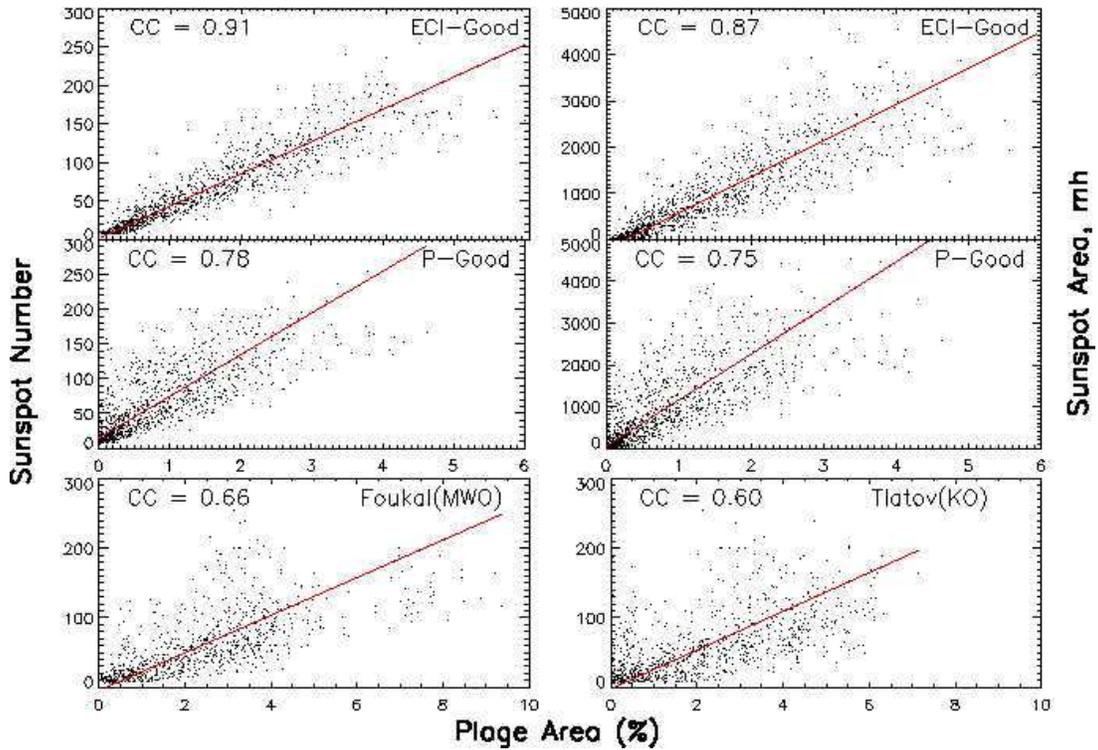}}
\caption{Two panels of the top row of the figure show the scatter plots between monthly plage areas determined using ``ECI-Good'' images and numbers (left) and areas (right) of the sunspot. Two panels in the middle row show the scatter plot between plage areas determined earlier using (P-images) and sunspot data. Two panels in the bottom row show a scatter plot of the plage area \citep{1996GeoRL..23.2169F} versus sunspot number (left) and plage area \citep{2009SoPh..255..239T} versus sunspot area (right). The value of the correlation coefficient is indicated in each panel. It may be noted that the scale of plage area along the x-axis is different for the bottom-row panels than top and middle-row panels.} 
\label{fig:4a}
\end{figure}

\begin{figure}[h]
\centerline{\includegraphics[width=0.8\textwidth]{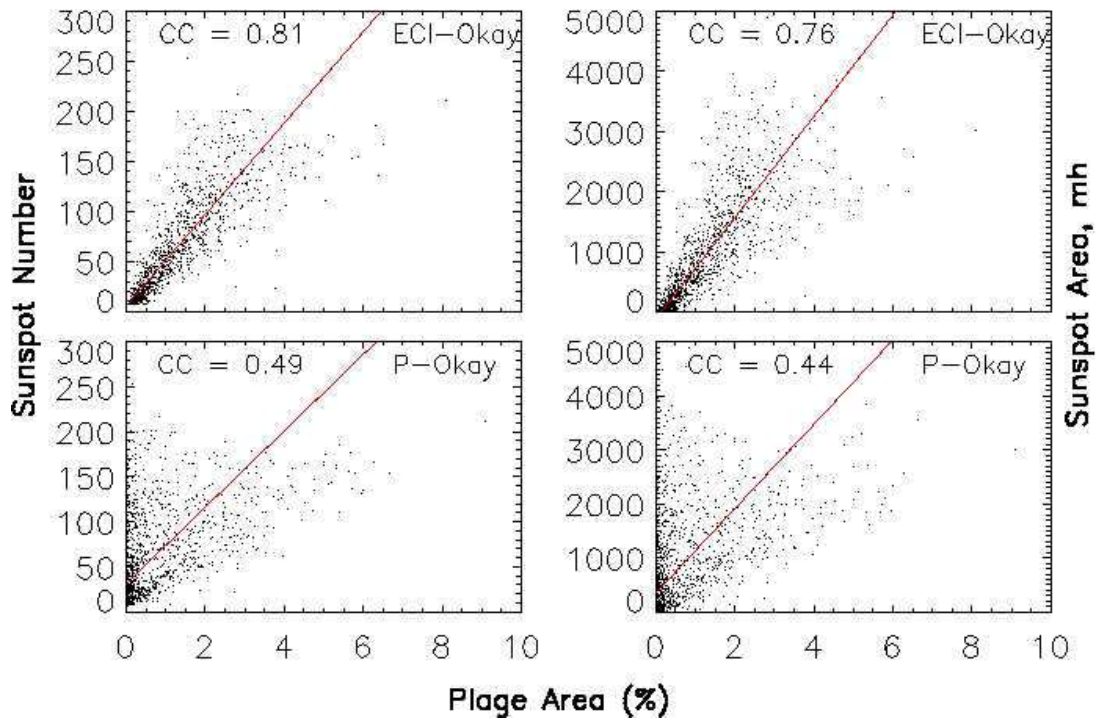}}
\caption{Two panels of the top-row of the figure show the scatter plots between monthly plage areas determined using ``ECI-Okay'' images and numbers (left) and areas (right) of sunspots. Two panels in the bottom row show the scatter plot between plage areas determined earlier using P-Okay images \citep{2019SoPh..294..131P} and sunspot data.} 
\label{fig:4b}
\end{figure}

The monthly averaged plage areas determined using the MWO,  KO and some other data are available \citep{1996GeoRL..23.2169F,2009SoPh..255..239T} for comparison for this period. Therefore, we compare sunspot numbers and plage areas on a monthly average basis to determine the most accurate methodology to analyze the historical data. In the top row of Figure~\ref{fig:4a}, we plot the monthly averaged plage areas determined using ECI-Good data versus WDC-SILSO sunspot number and RGO sunspot area for the period 1905 -- 2004 (http://www.sidc.be/silso/datafiles). 

The middle row of this figure shows the plage areas determined from the P-images before applying the equal contrast technique versus sunspot number (left) and area (right). The number of data points in all four plots is the same. In the bottom row of Figure~\ref{fig:4a}, we show the scatter plot of sunspot number versus plage area identified by Foukal (1996) using MWO data for the period 1915 -- 1985 only, (ftp://ftp.ngdc.noaa.gov/STP/SOLAR\_DATA/SOLAR\_CALCIUM/DATA/Mt\_Wilson/) in the left-side panel and by \citet{2009SoPh..255..239T} using KO data in the right-side panel for the period 1907 -- 1999. The values of correlation coefficients are indicated in each panel. Similarly, two panels in the top row of Figure~\ref{fig:4b} show the scatter plots of the plage area determined from ECI-Okay images versus monthly average sunspot numbers (left) and sunspot area (right). Two panels in the bottom row show the same for P-Okay images. The values of the correlation coefficient are indicated in the panels. 

\begin{table}[h]
\centering
%\begin{center}
\caption{Correlation coefficients for the monthly averaged data for comparison. In the Table ssn and ssa represent sunspot number and sunspots area respectively.}
\begin{tabular}{|c| c| c|}
\hline
Data Type & Plage area v/s ssn & Plage area v/s ssa  \\
ECI-Good (KO) & 0.91 & 0.87  \\
ECI-Okay (KO) & 0.81 & 0.76 \\
ECI-Combined (KO) & 0.9 & 0.82  \\
P-Good (KO) & 0.78 & 0.75 \\
P-Okay (KO) & 0.49 & 0.44 \\
\citet{1996GeoRL..23.2169F}(MWO) & 0.66 & -- \\
\citet{2009SoPh..255..239T}(KO) & 0.6 & --\\
\hline
\end{tabular}
\label{tab:2}
\end{table}

%\end{center}

The plots in the middle row (Figure~\ref{fig:4a}) of monthly averaged data (P-Good) show a large scatter with correlation coefficients around 0.75. But, the plage area obtained from the ECI-images shows that the correlation coefficient improved significantly to a value of $\sim$~0.9. Similarly, the plots in Figure~\ref{fig:4b} for the ``ECI-Okay'' images indicate a decrease in the scatter and large improvement in the correlation between plage areas and sunspot parameter. The correlation coefficient improves from a value of $\sim$~0.45 (P-Okay) to $\sim$~0.8 for ECI-Okay images. In Table~\ref{tab:2}, we show the values of correlation coefficients for MWO and KO data sets derived by different authors for easy comparison. The obtained values of correlation coefficients are significantly larger for ECI-Good and ECI-combined data sets compared to others. It may be noted that the correlation coefficients are better for the ``ECI-Good'' data as compared to that for the ``ECI-Okay'' data, even in the case of equal contrast images. The scatter plots between sunspot data and the plage areas indicate a much better correlation after making contrast equal, of all the images for the ``P-Good'' data and the ``P-Okay'' data. From the scatter plots, we learn that it is possible to combine the ``ECI-Good'' and ``ECI-Okay'' after applying the equal contrast technique. This will reduce the data gaps in the time series. The good values of correlation coefficients between sunspot data and plage area obtained using ECI-images compared to the other procedures adopted earlier to analyze Ca-K images of the historical data \citep{2019SoPh..294..131P, 1996GeoRL..23.2169F, 2009SoPh..255..239T} indicate that the difference in the correlation coefficients is likely due to a combination of the different techniques, spectral bandwidth, and spatial resolution between the two databases.
 
\begin{figure}[h]
\centerline{\includegraphics[width=0.8\textwidth]{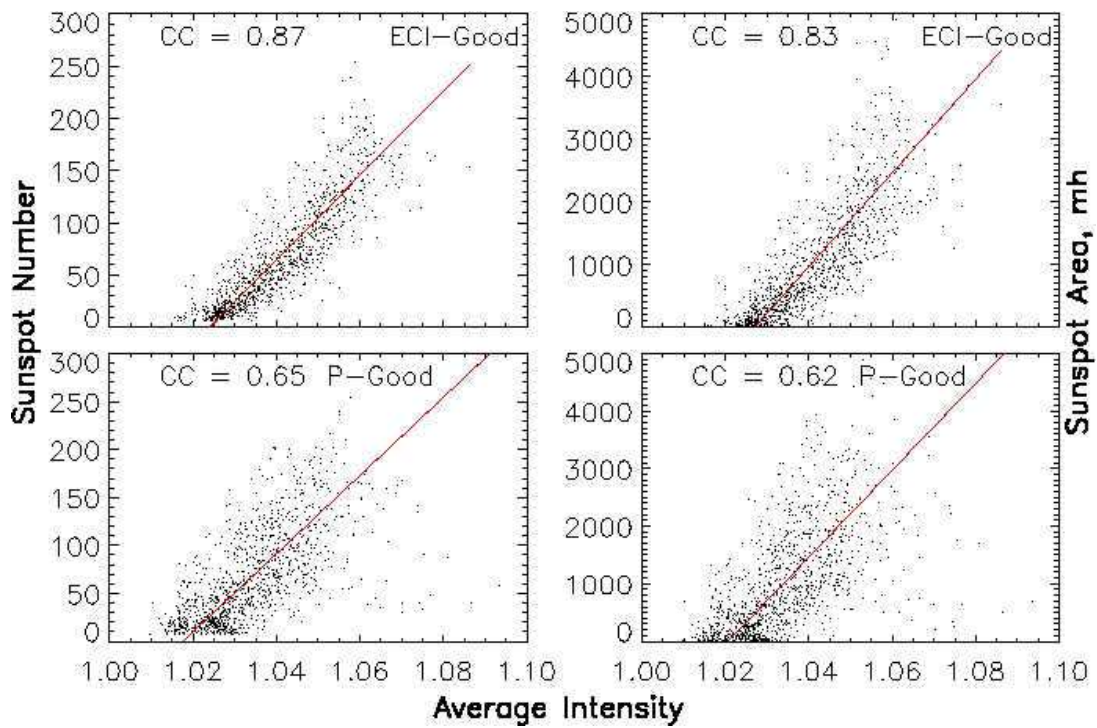}}
\caption{Two panels of the top row of the figure show the scatter plots between monthly averaged intensity (Ca-K index) determined for ``ECI-Good'' images and sunspot numbers (left) and areas (right). Two panels in the bottom row show the scatter plot between averaged intensity (Ca-K index) determined earlier using P-Good \citep{2019SoPh..294..131P} and sunspot data.} 
\label{fig:5a}
\end{figure}

\begin{figure}[h]
\centerline{\includegraphics[width=0.8\textwidth]{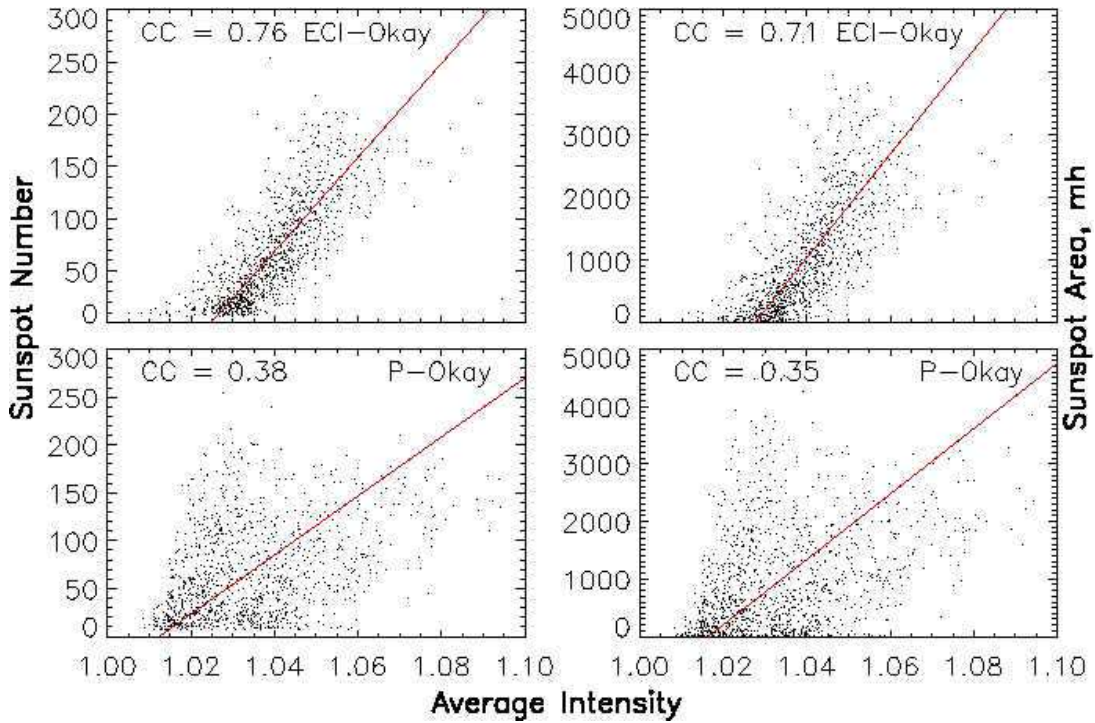}}
\caption{Two panels of the top row of the figure show the scatter plots between monthly averaged intensity (Ca-K index) determined for ``ECI-Okay'' images and sunspot numbers (left) and areas (right). Two panels in the bottom row show the scatter plot between averaged intensity (Ca-K index) determined earlier for P-Okay images \citep{2019SoPh..294..131P} and sunspot data.}
\label{fig:5b}
\end{figure}

Further, we have computed the average intensity (Ca-K index) over the whole of the disk image as done in our earlier paper \citep{2019SoPh..294..131P}. In the top-row of Figure~\ref{fig:5a}, we show the scatter plot of the full-disk intensity of the ``ECI-Good'' images versus sunspot numbers (left-panel) and sunspot area (right) for monthly averaged data . The bottom-row of this figure shows a similar scatter plot for the P-Good images. The scatter plots between the monthly averaged sunspot data and the averaged intensity of the ``ECI-Good'' images indicate an excellent correlation with a value of 0.85 and confidence level $>$ 99\% compared to the correlation coefficient of  0.65 for the plots in the bottom row of P-Good images. The plots pertaining to the recent study indicate less scatter in the data than earlier work \citep{2019SoPh..294..131P}. We show the linear least square fits for all the plots. There appears a polynomial relation between sunspot numbers and average Ca-K intensity for the ECI-images. But linear fit to the data points is more satisfactory as compared to 2 and 3-degree polynomial fits. The plots between sunspot data and the average intensity of the full-disk image for the ECI-Okay and P-Okay images, indicate a similar behavior, as shown in Figure~\ref{fig:5b}. The correlation coefficients of about 0.7 for ECI-Okay and 0.35 for P-Okay indicate more improvement in the correlation between monthly averaged Ca-K index and sunspot parameters in the present methodology. This is because a many low and high contrast images in this data set have been converted to images of the same contrast. It may also be noted that the correlation between the monthly averaged Ca-K index and sunspot data is better for the ``Good'' data as compared to the ``Okay'' data in both the cases with and without, conversion of images to equal contrast.

\subsection{Variation of Ca-K Area Index with Time using ECI-images}
In \citet{2019SoPh..294..131P}, the plage areas determined from the ``P-Good'', and ``P-Okay'' data exhibit solar cycle variations, and their amplitude is in general agreement with the sunspot data used with a 12-months running average. That time the procedure adopted to analyze the historic data provided reliable results till 1984 only, and the amplitude of variation for a couple of solar cycles differed with those of sunspot cycles. Now, we have developed the methodology of the Equal contrast technique to analyze that data further. We have shown that ECI-Okay data become comparable with the ECI-Good data. But to make a detailed comparison, we analyze these two series separately before combining the whole data.

\begin{figure}[h]
\centerline{\includegraphics[width=0.8\textwidth]{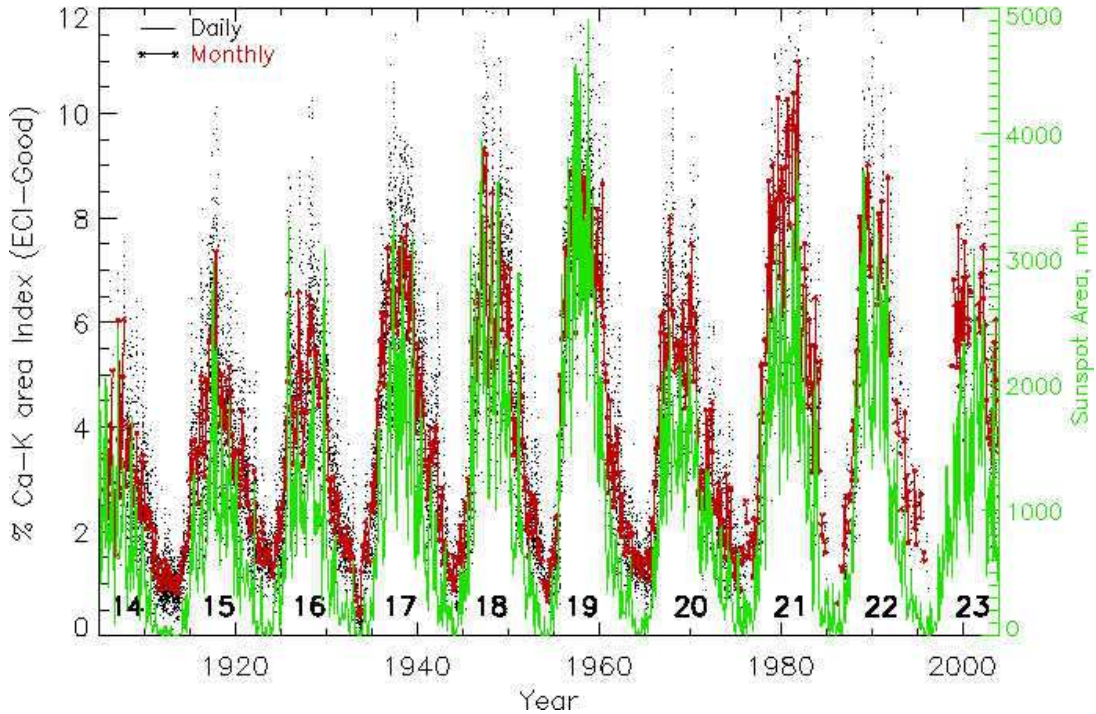}}
\caption{The black dots in the figure show the percentage of daily Ca-K area index as a function of time for ``ECI-Good'' images. The red curve indicates the monthly average of the Ca-K area index, and the green curve shows the monthly average of sunspot area for the period 1905 -- 2004. The cycle numbers are also indicated in the figure.}
\label{fig:6a}
\end{figure}

\begin{figure}[h]
\centerline{\includegraphics[width=0.8\textwidth]{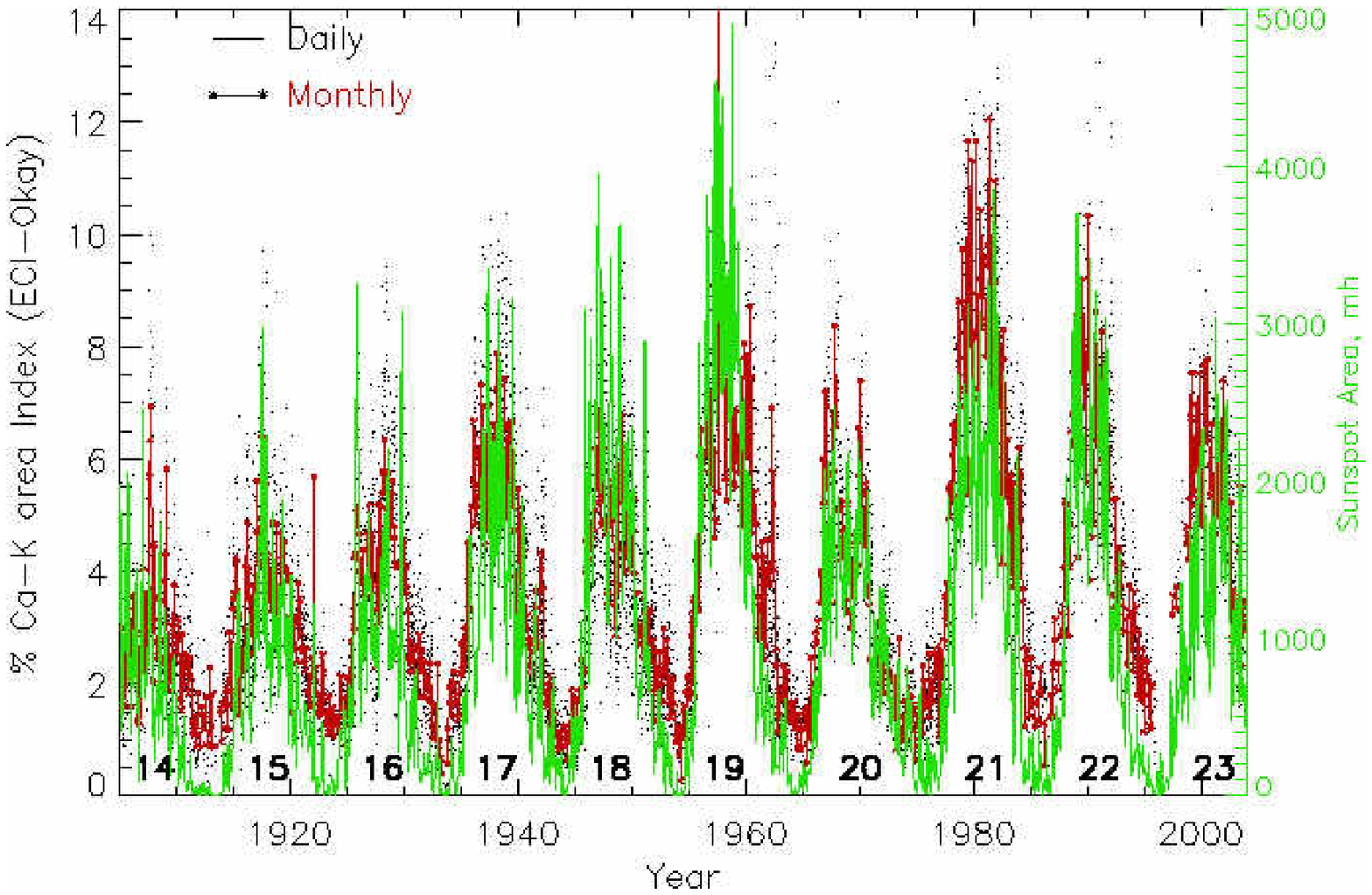}}
\caption{The black dots in the figure show the percentage of daily Ca-K area index as a function of time for ``ECI-Okay'' images. The red curve indicates the monthly average of the Ca-K area index, and green the monthly average of sunspot area for the period 1906 to 2007. The cycle number is also indicated in the figure.}
\label{fig:6b}
\end{figure}

After making all the images of equal contrast, we again examined the intensity threshold values to identify the plages and network features. After several experiments, we found the earlier values of intensity threshold determined by \citet{2019SoPh..294..131P} still holds good. This is because average values of FWHM of intensity distributions for the P-Good and ECI-Good data are similar. We have defined plages with intensity contrast $>$ 1.30 and consecutive area $>$ 1000~pixel$^{2}$, equivalent to about 0.2~arcmin$^{2}$. In addition, we define intensity $>$ 1.30 and consecutive area $>$ 4 pixels as enhanced network (EN) and intensity $>$1.20 and $<$ 1.30 with consecutive area $>$ 4 pixels as active network (AN).
The regions with intensity $>$~1.1 and $<$~1.2 with consecutive area $>$~4 pixels are treated as quiet network (QN). To compare the Ca-K and sunspot data, we computed the sum of the percentage of plage, EN, and AN (hereafter called Ca-K area index) from daily data. The difference between the total active area and Ca-K area index is that the total active area includes the QN area, also. We compare both the sum of plages, EN, and AN area variation (Ca-K area index mostly related with activity) and total active area (including QN) with the sunspot activity. The QN may be related with the global characteristic of sun rather than sunspot activity . In Figure~\ref{fig:6a}, we plot the Ca-K area index represented by black dots, monthly averages of Ca-K area index indicated by the red curve, and monthly averages of sunspot areas shown  in green  for the  ``ECI-Good'' time series for the period of 1905 -- 2004. The daily Ca-K area index shows that the scatter in the derived Ca-K area index is very less as compared to similar plots by \citet{2019SoPh..294..131P} and others. In our earlier paper \citep{2019SoPh..294..131P}, we plotted the data for the period of 1907 -- 1984 as there was a large scatter in the determined values of plage areas because of varying sky conditions and change of photographic emulsion. But after equal contrast technique applied to the images, it has become possible to determine the plage areas reliably for 1985 -- 2007 data, also. The respective solar cycle numbers are also written in the figure. The plot indicates an excellent agreement between the maximum amplitudes of solar cycles in the Ca-K area index with sunspot areas except for solar cycle number 21 (around the year of 1980). The difference may be due to the availability of a fewer number of images per year after the year 1980. Generally, the red curve amplitudes appear more than those of green curves due to the selected scales of the Ca-K area index and sunspot areas.  

Figure~\ref{fig:6b} shows the daily Ca-K area index, monthly averages of Ca-K area index, and the monthly average of sunspot areas from 1905 to 2007 by black dots, red and green curves, respectively, for the ``ECI-Okay'' data. The period of ``ECI-Good'' and ``ECI-Okay'' data differs because there are no ``P-Good'' images available from 2005 to 2007. The derived values of the Ca-K area index appear more reliable as these show much less scatter as compared to earlier studies by \citet{2019SoPh..294..131P} and others. Generally, the monthly average Ca-K area index shows good correlations with sunspot area values. But, the amplitude of monthly averages of the Ca-K area index for the ``ECI-Okay'' data differs from those for the ``ECI-Good'' data, especially for cycle numbers 18 and 19. 

\subsection{Variation of Ca-K intensity index with time using ECI-images}
\begin{figure}[h]
\centerline{\includegraphics[width=0.8\textwidth]{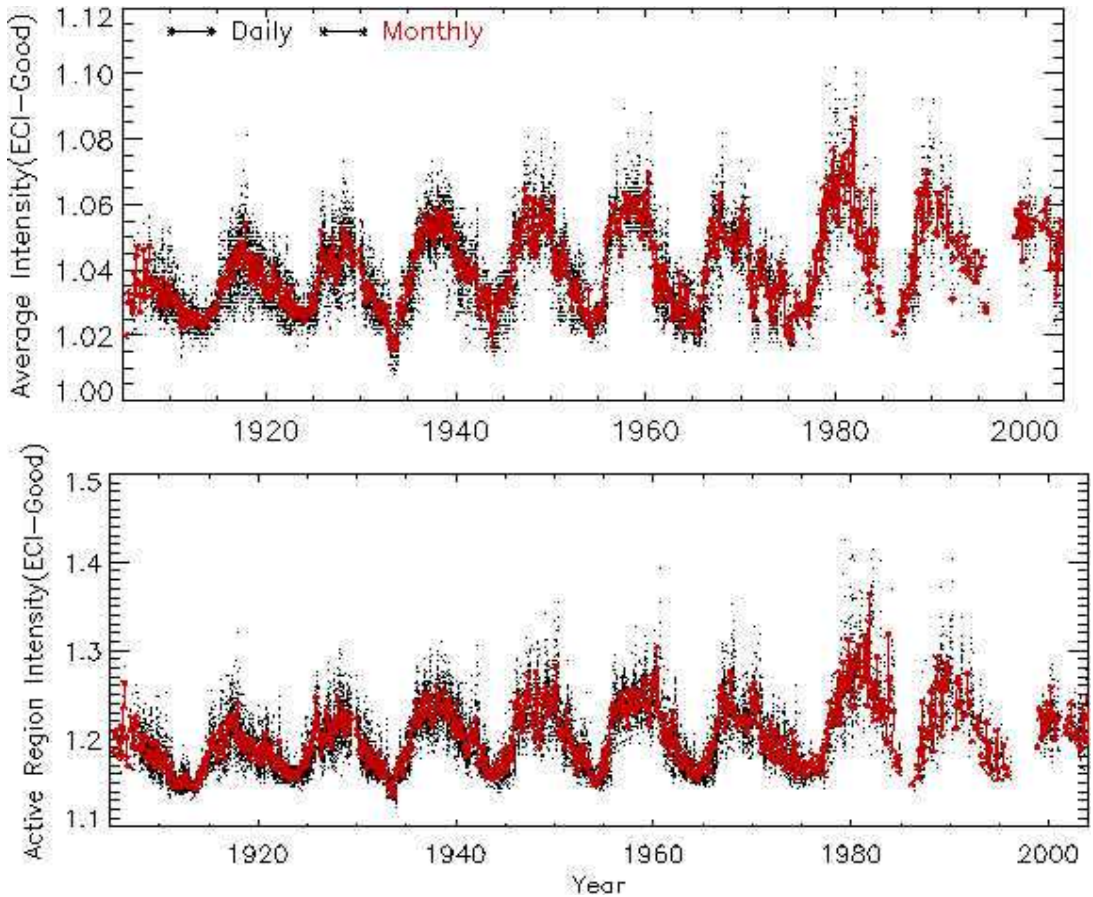}}
\caption{The upper panel of the figure shows the average intensity over the solar image computed after normalizing the peak of the intensity distribution to unity on a daily basis whenever observed by black dots for the data termed as ``ECI-Good''.  The red curve indicates the monthly averaged intensity of the Ca-K images. The bottom panel shows the average active region intensity on a daily basis shown by black dots and the monthly average by the red curve.}
\label{fig:7a}
\end{figure}

\begin{figure}[h]
\centerline{\includegraphics[width=0.8\textwidth]{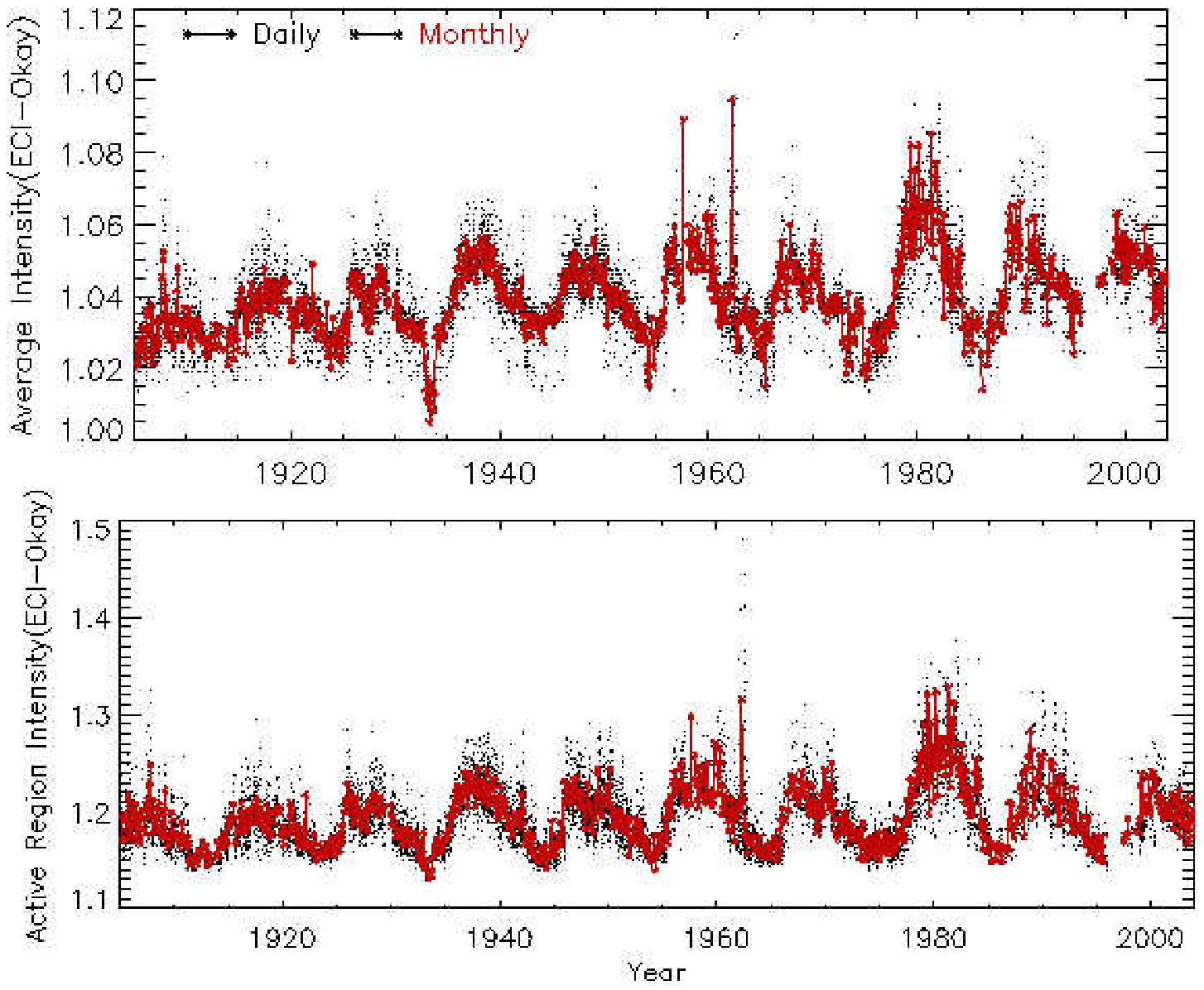}}
\caption{The upper panel of the figure shows the average intensity computed after normalizing the peak of the intensity distribution to unity over the image on a daily basis whenever observed by black dots for the data termed as ``ECI-Okay''. The monthly averaged intensity of the Ca-K images is indicated by the red curve. The bottom panel shows the average active region intensity on a daily basis by black dots and the monthly average by the red curve.}
\label{fig:7b}
\end{figure}

Black dots in the upper panel of Figure~\ref{fig:7a} show average intensity over the whole solar image on a daily basis and monthly average intensity (red curve) for the period of 1905 -- 2004 for the ``ECI-Good'' time series.  We have also computed the average intensity of the active region with pixels having intensity contrast $>$1.1 and is shown in the bottom panel of the same figure. The full-disk intensity and active region intensity for the ``ECI-Good'' data show that the amplitude variation increases from solar cycle 14 to 19, similar to that of sunspots data shown in Figures~\ref{fig:6a} and \ref{fig:6b}. The determined intensities vary smoothly with the phase of the solar cycle. This confirms that by fine-tuning the contrast of the images, the data has become uniform in quality. We show the same parameters for the ``ECI-Okay'' images for the period of 1905 -- 2007 in two panels of Figure ~\ref{fig:7b} that indicates results similar to those of ``ECI-Good'' data but with some scatter. This is probably due to the quality of images but not due to the contrast of images or could be  because of some other reason.

\subsection{Variation of Ca-K networks area and total active area with time using ECI-images}

\begin{figure}[h]
\centerline{\includegraphics[width=0.8\textwidth]{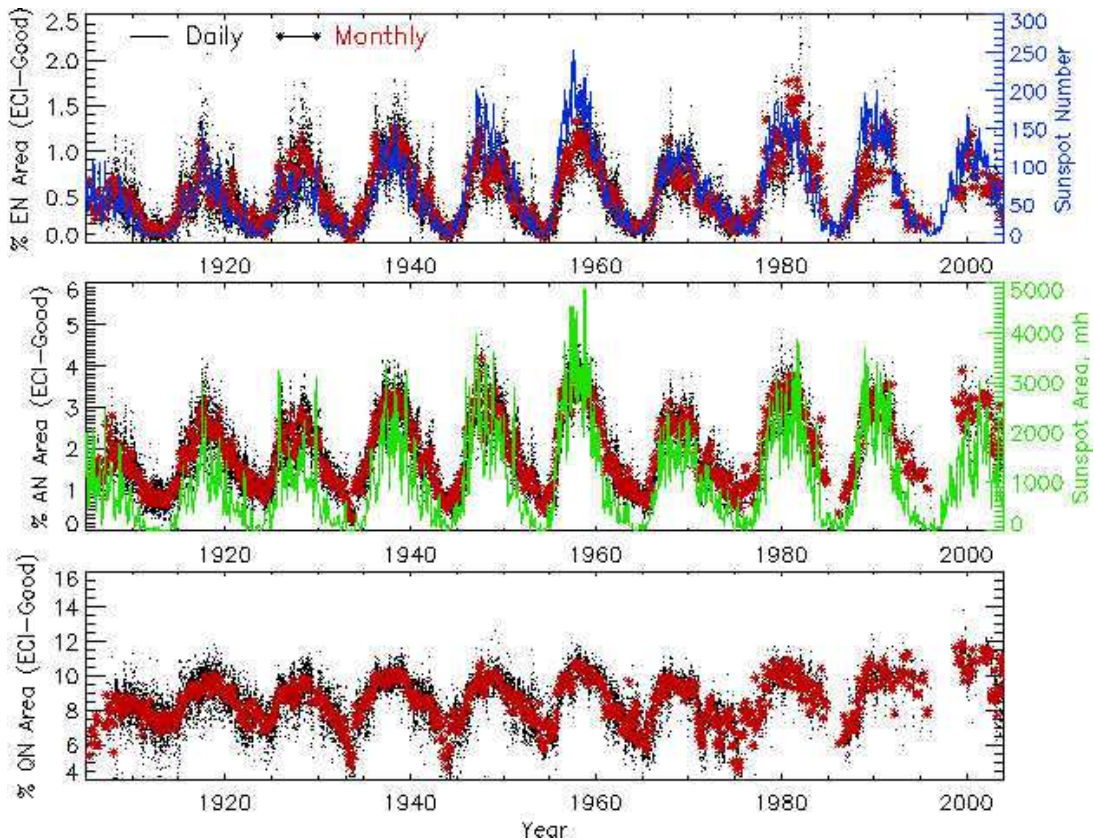}}
\caption{The top panel of the figure shows \% of enhanced network area (EN), monthly averages, and monthly averages of sunspot numbers by black dots, red and blue curves, respectively, as a function of time for the ``ECI-Good'' data. The percentage of active network area (AN), monthly averages of AN, and monthly averages of sunspot areas are shown in the middle panel. The bottom panel shows the variations of the quiet network area (QN) in percentage.}
\label{fig:8a}
\end{figure}

\begin{figure}[h]
\centerline{\includegraphics[width=0.8\textwidth]{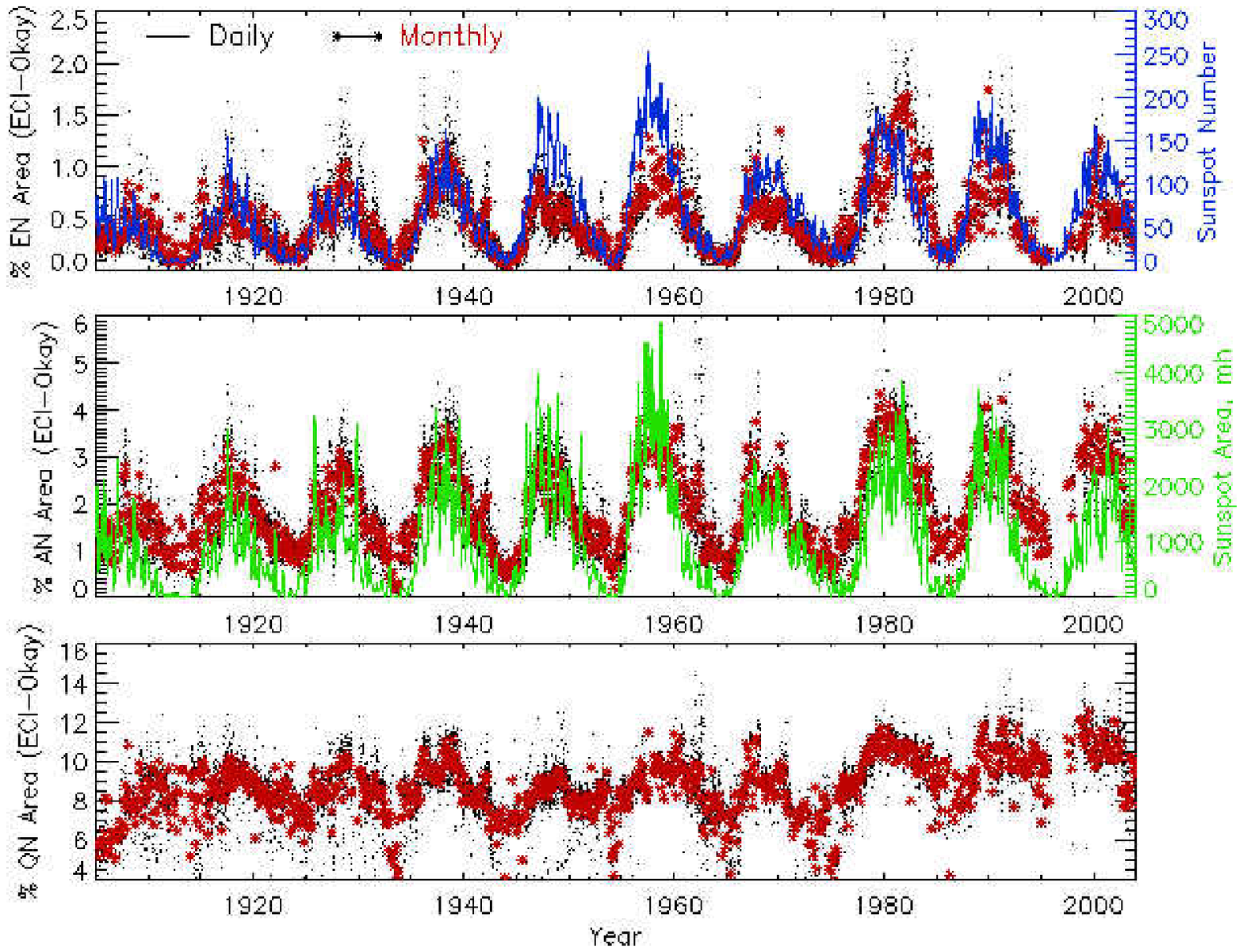}}
\caption{Same as that for Figure~\ref{fig:8a} but for the ``ECI-Okay'' data.}
\label{fig:8b}
\end{figure}

The top panel of Figure~\ref{fig:8a} shows the time-variation of enhanced network area (EN), representing the decaying plage regions as a function of time on a daily basis shown by black dots and monthly averaged data displayed by the red curve for the ``ECI-Good'' time series. The monthly averages of sunspot numbers are over-plotted for the comparison. The middle panel of the figure shows the variation of active network area (AN) on a daily basis (black dots), monthly averages (red curve), and the monthly averages of sunspot area (green curve). The bottom panel indicates the variation of quiet network area (QN) on a daily basis (black dots), month averages (red curve). The plots of EN and AN show that amplitudes of variations agree well with those of sunspot data, whereas the amplitude of variations for QN remains more or less the same throughout all the solar cycles, 14 -- 23. Three panels of Figure~\ref{fig:8b} show variations of EN, AN, and QN on daily and monthly average basis for the data termed as ``ECI-Okay'' as a function of time. The plots indicate variations in ``ECI-Okay'' data similar to those for the ``ECI-Good'' data. It may be noted that there appears some scatter in ``ECI-Okay'' data in addition to temporal variations.

\begin{figure}[h] 
\centerline{\includegraphics[width=0.8\textwidth]{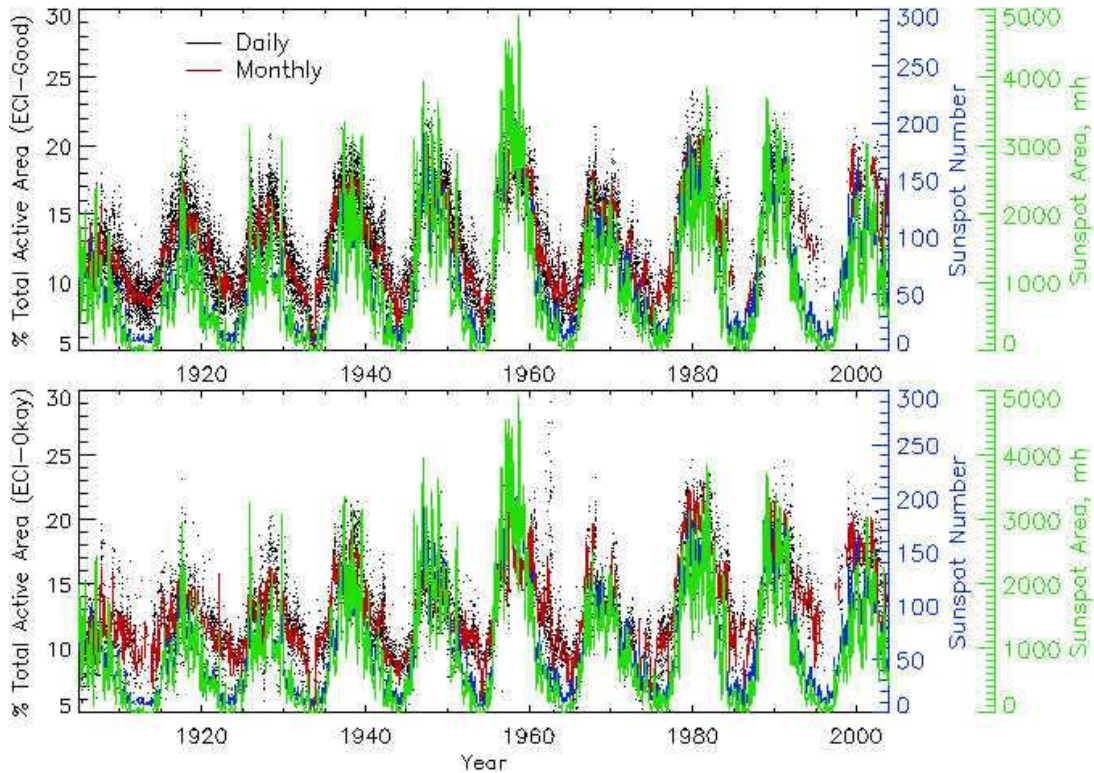}}  
\caption{The Figures's upper panel shows the percentage of the total active area of images on a daily basis shown as black dots and monthly averages are shown in the red for ``ECI-Good'' data as a function of time. We have over-plotted monthly averages of sunspot numbers and areas for comparison. The bottom panel shows the same for ``ECI-Okay'' data.}  
\label{fig:9}
\end{figure}

We have also computed the total active area in the images with intensity contrast of pixels $>$ 1.1 and consecutive area $>$ 4~pixels (3 arcsec$^{2}$). The upper panel of Figure~\ref{fig:9} shows the percentage of the total active area on daily basis (black dots), monthly averages (red curve), monthly averages of sunspot numbers (blue), and monthly averages of sunspot area in green as a function of time for the ``ECI-Good'' data. The amplitude of variations in all the parameters agrees well with each other for all the solar cycles. This figure’s bottom panel indicates the same parameters for ``ECI-Okay'' data for the period of 1905 --2007. The plots show the results similar to those for ECI-Good data but with some scatter due to the quality of some of the images in the ECI-Okay time series.

\subsection{Variations with combined ``ECI-Good'' and ``ECI-Okay'' data}

\begin{figure} [h]
\centerline{\includegraphics[width=0.8\textwidth]{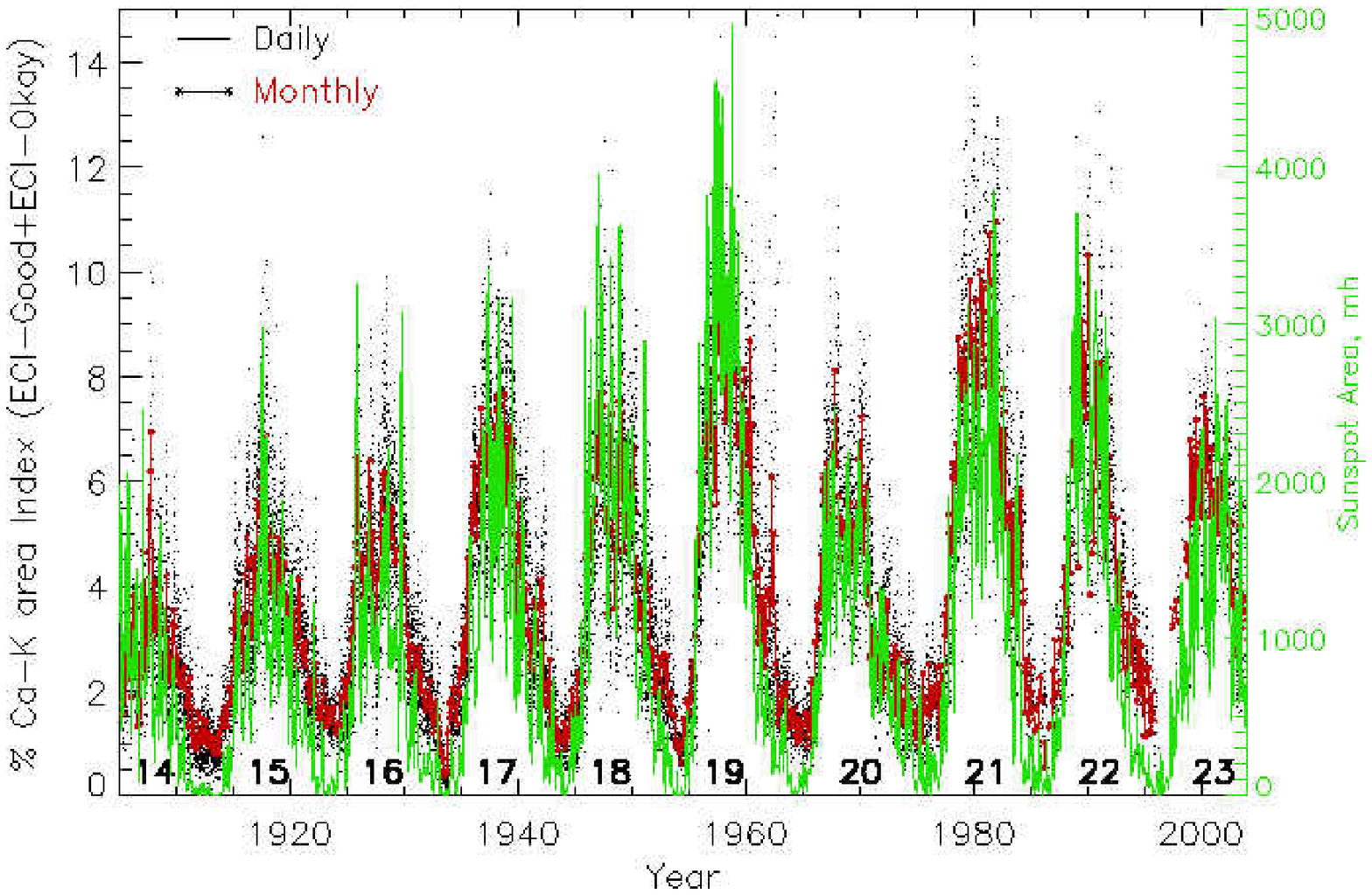}}  
\caption{The black dots in the figure show the daily Ca-K area index as a function of time for the whole data of ``ECI-Good'' and ``ECI-Okay'' time series. The red curve indicates the monthly average of the Ca-K area index, and green the monthly average of sunspot area from  1905 to 2007. The cycle number is also indicated in the figure.}
\label{fig:10}
\end{figure}

Further, we have combined both the ``ECI-Good'' and ``ECI-Okay'' times series to study the variations due to all the Ca-K line images obtained at Kodaikanal observatory. In Figure~\ref{fig:10}, we show the Ca-K area index on a daily basis by black dots, its monthly averages by the red curve, and monthly averages of sunspot areas by a green curve as a function of time for the period 1905 -- 2007. The plots indicate a good correlation between the amplitudes of the Ca-K area index and sunspots areas for all the solar cycles. It may be noted that the inclusion of ``ECI-Okay'' images has decreased the amplitude of the Ca-K area index for the solar cycle numbers 18 and 19 as compared to those for the ``ECI-Good'' data.

\begin{figure} [h]
\centerline{\includegraphics[width=0.8\textwidth]{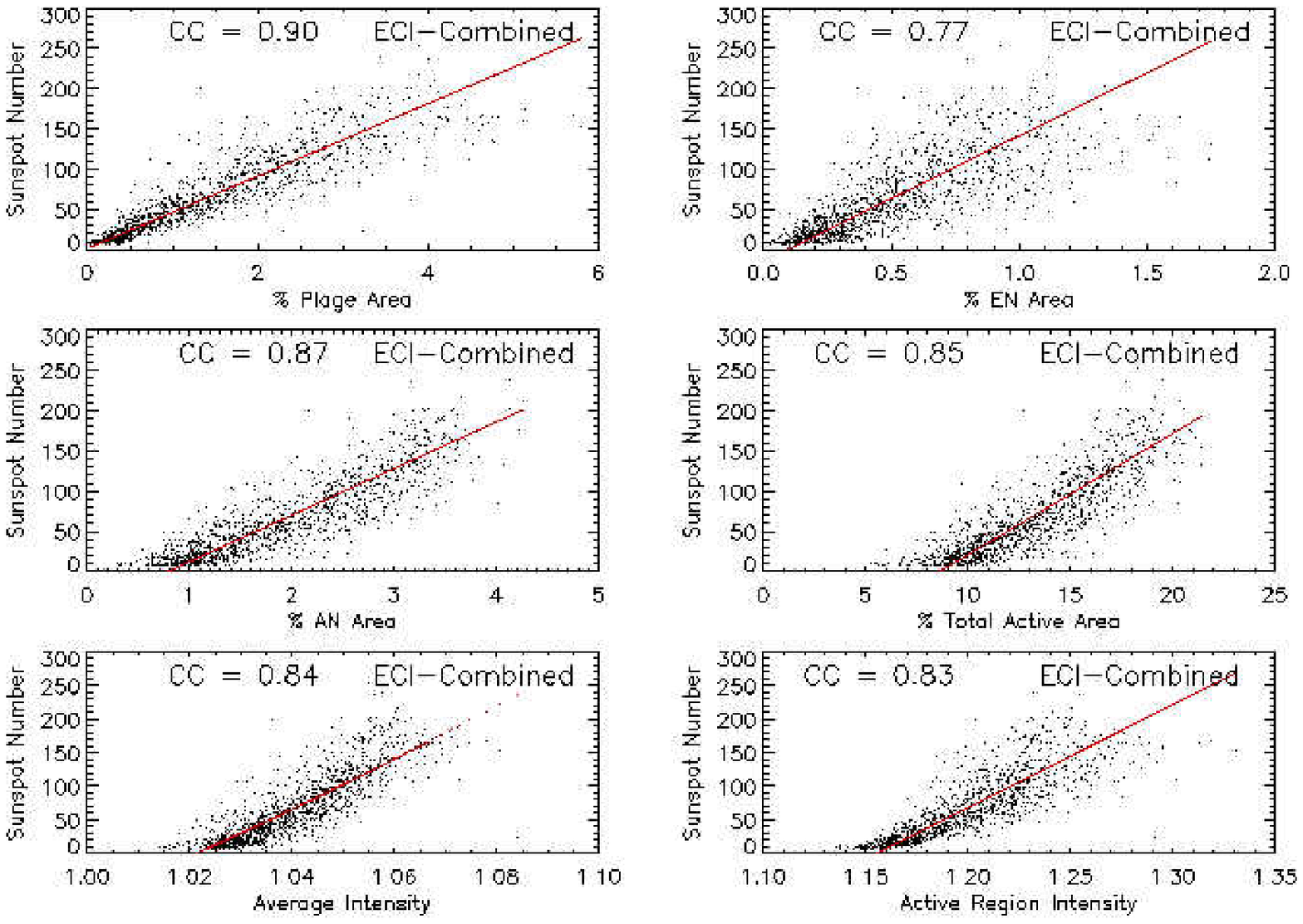}}  
\caption{Top two panels of the figure show the scatter plots of sunspot numbers versus \% of Ca-K plage area (left) and \% of EN (right) considering all the data of ``ECI-Good'' and ``E-Okay'' time series on a monthly average basis. The middle two panels show the scatter plot of sunspot number versus \% of AN (left) and \% of total active area (right). The bottom two panels show average intensity (left) and total active intensity (right). The value of the correlation coefficient is indicated in each panel.}
\label{fig:11}
\end{figure}

\begin{figure} [h]
\centerline{\includegraphics[width=0.8\textwidth]{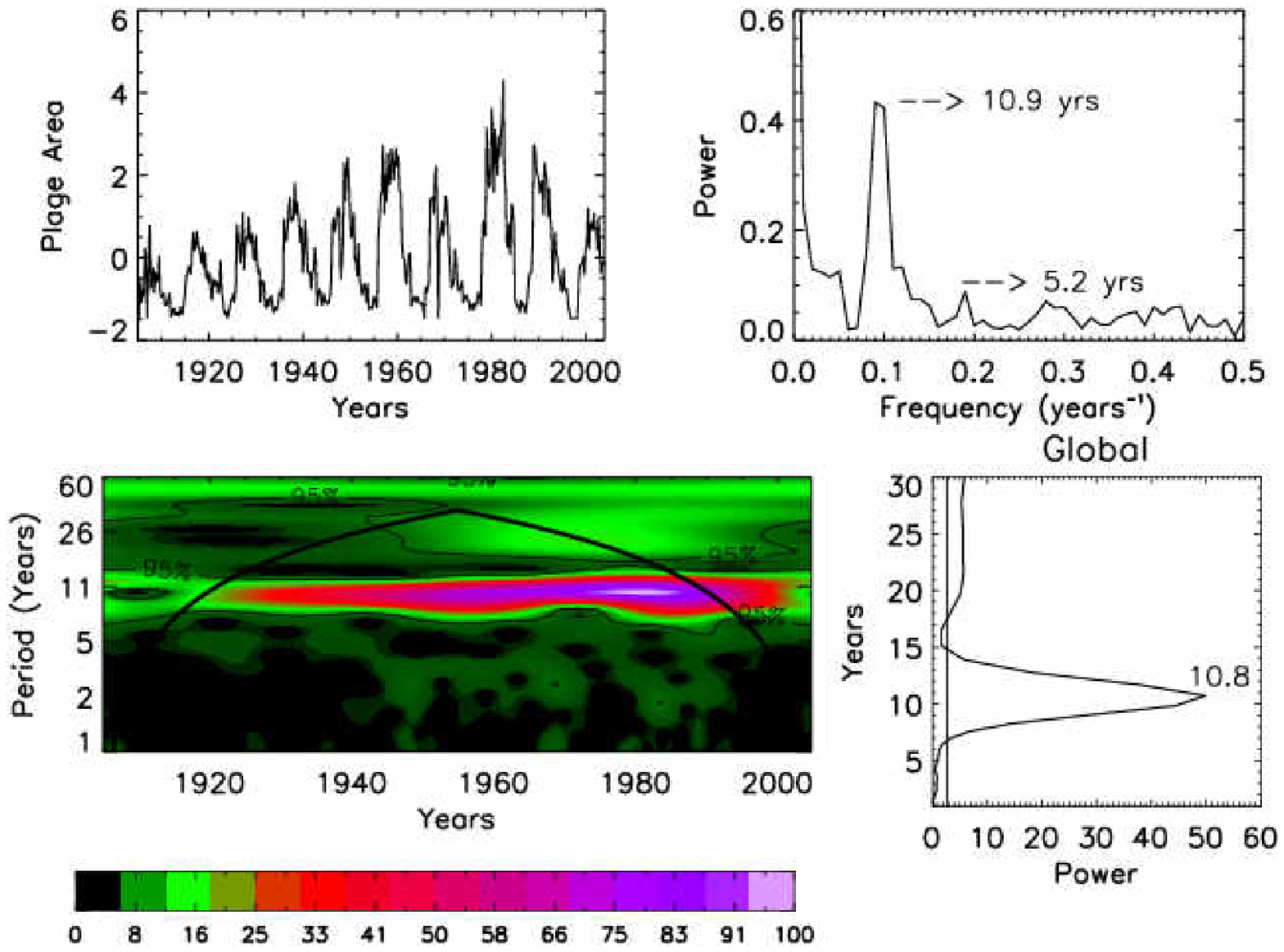}}  
\caption{The top left panel of the figure shows relative variations in the monthly plage areas of the combined ECI-images for the period of 1905 -- 2007 and the top-right panel shows the power spectrum. The bottom-left panel shows the wavelet power spectrum, and the right-side panel shows the global power spectrum of the data. The colour bar at the bottom indicates the relative power in the power spectrum.}
\label{fig:12}
\end{figure}

Six panels of Figure~\ref{fig:11} show the scatter plots between the monthly average of sunspot numbers and monthly averages of Ca-K plages, EN, AN, total active area, the average intensity of image, and active region intensity (average intensity of all areas with normalized intensity greater than 1.1). The plots for AN and total active area  appear to indicate polynomial relation but the linear fits were found to be better than polynomial fits. Generally, it can be stated that there is a linear relation between numbers and areas of sunspots and Ca-K line parameters. An excellent correlation between the monthly averaged sunspot numbers and various features visible in Ca-K line images indicates that this methodology permits us to study the long term systematic variations in small scale activity apart from the large scale activity. 

\subsection{Periodic behaviour in Ca-K feature}
\label{sec:7}
We have performed the wavelet analysis of monthly averaged data of plage, EN, AN, and QN areas for the combined data of ``ECI-Good'' and ``ECI-Okay'' times series to investigate the periodicities in variations of the large and small scale activity on the sun. The top row two panels of Figure~\ref{fig:12} show the relative variations in the plage areas as a function of time (left) and its power spectra (right). The bottom two panels show the wavelet power spectrum (left) and the global power spectrum (right). The color bar indicates the relative power in the wavelet power spectrum. The power spectrum and the global power spectrum indicates a strong periodicity around 10.8-years period and negligible power at periods $>$ 16 and $<$ 5 years. The existence of some quasi-periods $<$ 4 years and the 11-years period was observed in the earlier analysis. The wavelet analysis of the EN and AN indicates similar results as seen in Figures~\ref{fig:12a}, \ref{fig:12b} (Annexure~1). But the wavelet analysis QN area in Figure~\ref{fig:12c} (Annexure~1) shows some power at periods $>$ 16 and $<$ 4 years along with large power around 11-years. This data set is more uniform than earlier ones and will be subjected to more rigorous analysis to study the existence of quasi-periods.

\section{Discussions}
We have developed a new methodology to analyze the Ca-K line images to investigate long-term variations in the chromosphere's large and small scale activity. First, we made all the images of equal contrast. This is to compensate for the different photographic emulsions used for recording the images, different sky transparency and seeing conditions affecting the contrast of the images, changes in contrast due to centering of the Ca-K line on the exit slit of spectro-heliograph, different developing conditions, and chemicals used for the developing photographs over long periods. This methodology has also overcome the difficulty of intensity calibration of the time series as a significant part of the data did not have step wedge calibration. Even the step wedge calibration has some uncertainty due to the non-uniformity of the light source used at edges of photographic plates \citep{2014SoPh..289..137P}. 

The variations in the parameters of Ca-K line features obtained from the analysis of each image represent variations on the sun, free from the effects of  above mentioned observational parameters and sky conditions. Most of the papers (\citet{2019SoPh..294..131P, 2020A&A...639A..88C} and the references therein) have done the 12-months averages or running averages of the data to show the correlation between sunspots and Ca-K plage areas. We have shown the correlation between plage areas and sunspot data on daily and monthly mean basis. The percentage of plage area vary with the solar cycle phase smoothly  and their amplitude agrees well with the amplitude of sunspot data till 1975 for the data termed as ``ECI-Good''. After 1975 the amplitudes of plage areas differs with those of sunspots by a small amount, probably due to available data for a fewer number of days per year. Apart from the study of variation of large scale activity represented by plage areas, we have successfully investigated the variations in small scale activity represented by EN, AN, and QN for the first time for about 100 years. Only \citet{1998ApJ...496..998W} has defined the threshold values of intensity for EN and AN using some selected data from 1980 to 1996 to study their relationship with solar cycle phase. The amplitudes of solar cycle variations for EN and AN agree well with those of sunspot data. The amplitude of solar cycle variations for QN remains almost the same irrespective of the amplitude of sunspot data. It may be noted that the values of average monthly intensity include the intensity of a quiet background chromosphere. This data will be very useful for making the realistic models of solar cycle variations and will be made available to the interested scientist. 

The Ca-K line images obtained at various observatories with different spatial resolution and passband lead to different contrast. For combining the data of different observatories and types, the derived parameters have been compared with each other with large temporal averages and then up or downscaled one of the data \citep{2016SoPh..291.2967B}. But in this process, data on day to day basis remains uncorrected. In the present methodology, we first adjust each image's contrast and then identify the features, such as plages, EN, AN, and QN. Thus, we make the correction on a day to day basis. Hence, this method is likely to work for different spatial resolution and with different passband images within certain limits and will help in combining the data to yield better results.  

Most of the solar cycle variation models are based on the observed changes in large scale magnetic activity. Some small scale magnetic field measurements have been done and analysed to study solar cycle variations \citep{2011ASPC..437..167S}. Using the full-disk observations of solar magnetograms from the Michelson Doppler Imager \citep{1995SoPh..162..129S} instrument, \citet{2012ApJ...745...39J} have found two components of varying small scale fields. One of the components whose magnetic flux is smaller than 32$\times$10$^{18}$~Mx exhibits cyclic variation in anti-phase with the sunspot cycle. Another one with the flux between (4.3 -- 38)$\times$10$^{19}$~Mx correlated with the solar cycle. It is possible to observe the magnetic fields at 0.1$^{\prime\prime}$ or better, but with a smaller field-of-view \citep{2015ApJ...806..174J}. The measurement of the magnetic fields was also made with an accuracy better than 5~G. With the results obtained from the accurate analysis of 100~years of Ca-K images of the sun, it will become possible to study and understand the small scale network and internetwork magnetic fields and its variations over the solar cycle better. 

\section{Conclusion}

We have developed a new methodology of the ``Equal Contrast Technique'' to analyze the Ca-K line's photographic images for long periods under different environmental conditions, uniformly. This procedure will help to analyze similar data sets obtained at several observatories with different instruments, uniformly. Then combine all the data to form a long time series with fewer gaps in the data for further studies. This technique helps to determine the variations in large scale and small scale activity with very high accuracy and reliability for the good quality data. It also helps to minimize the errors in the low-quality data. It will be possible to extend the study of large and small scale magnetic activity on the sun back in time as the activity observed in the Ca-K line is related to the magnetic field of the sun.

\section{Acknowledgments}
We thank the referee, Alexei Pevtsov and Kiran Jain for providing valuable comments on the paper. Jagdev Singh designed and developed the digitizer units with the help of P. U. Kamath and F. Gabriel.  He trained the team to digitize the data and supervised the process to digitize the Ca-K line images. We thank all the observers at Kodaikanal Observatory who made the observations since 1904, kept the data in suitable environmental conditions, and digitized all the images. The daily sunspot data used here was from the SILSO sunspot numbers (WDC-SILSO, Royal Observatory of Belgium, Brussels). The monthly sunspot data was taken from http://www.sidc.be/silso/datafiles. The monthly Ca-K index was dowloaded from ftp://ftp.ngdc.noaa.gov/STP/SOLAR DATA/SOLAR CALCIUM/DATA/Mt Wilson/.

\section{Annexure 1}

\begin{figure}[!h] 
\centerline{\includegraphics[width=0.8\textwidth]{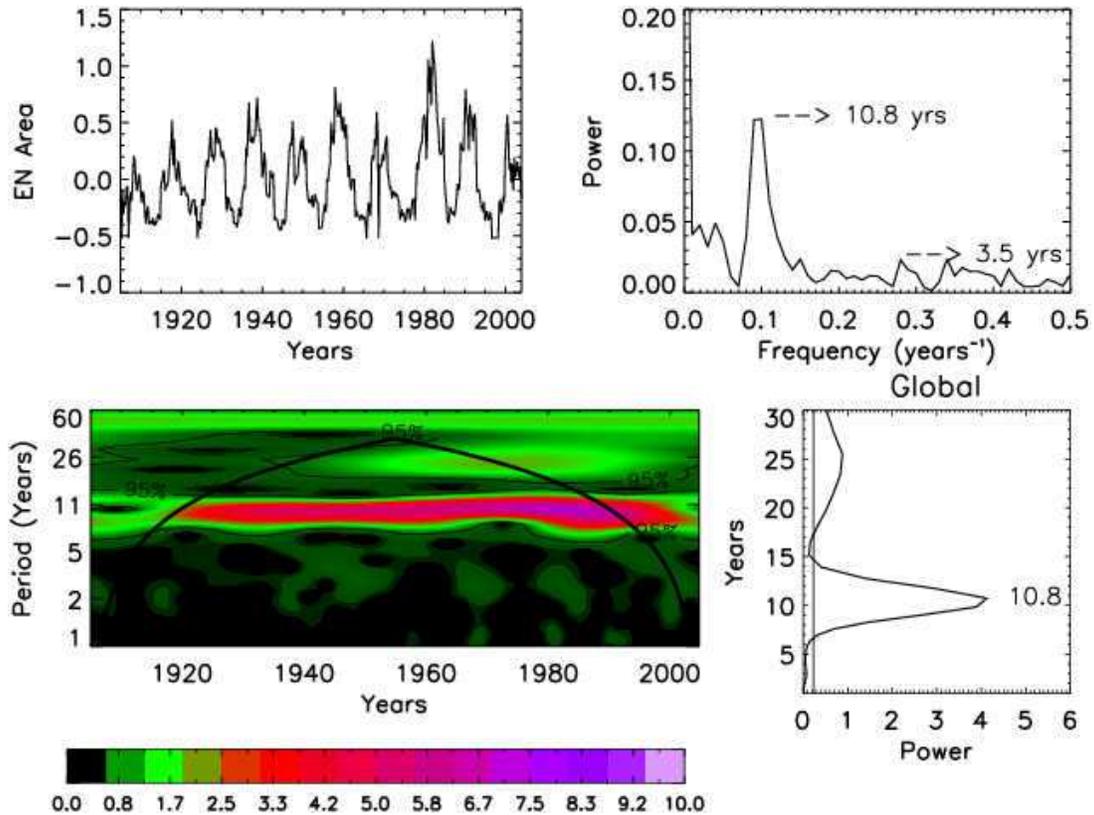}}  
\caption{The top left panel of the figure shows relative variations in the monthly EN areas of the combined ECI-images for 1905 -- 2007 and the top-right panel shows the power spectrum. The bottom-left panel shows the wavelet power spectrum, and the right-side panel shows the global power spectrum of the data. The colour bar at the bottom indicates the relative power in the power spectrum.}
\label{fig:12a}
\end{figure}

\begin{figure}[!h] 
\centerline{\includegraphics[width=0.8\textwidth]{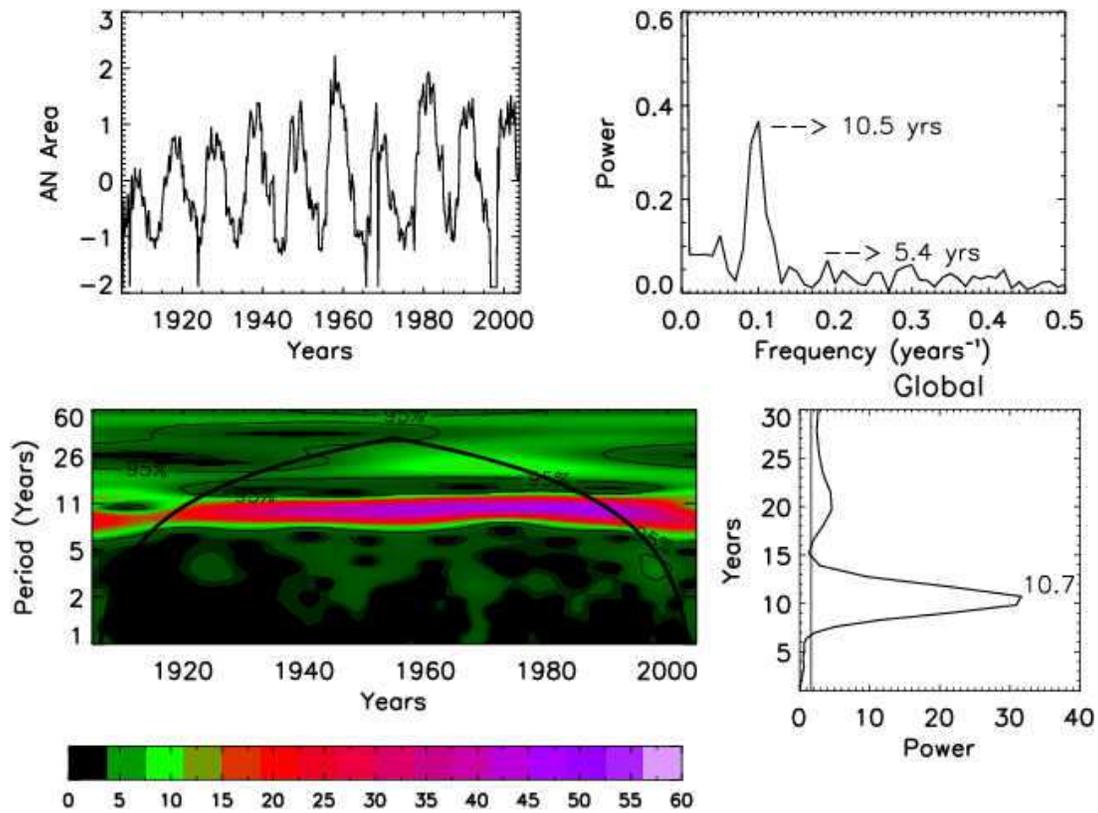}}  
\caption{Same as Figure~\ref{fig:12a} but for active network (AN).}
\label{fig:12b}
\end{figure}

\begin{figure}[!h] 
\centerline{\includegraphics[width=0.8\textwidth]{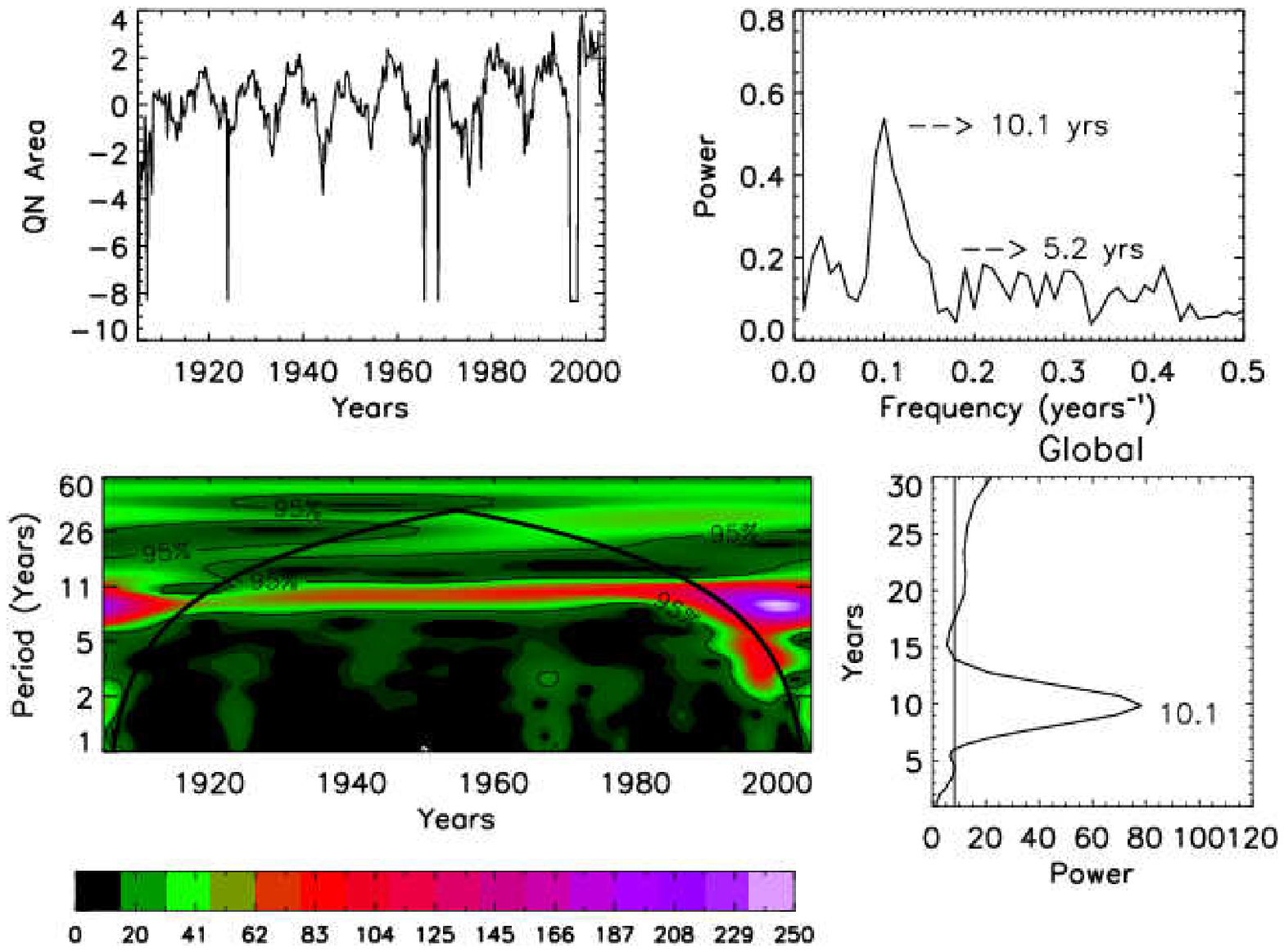}}  
\caption{Same as Figure~\ref{fig:12a} but for quiet network (QN).}
\label{fig:12c}
\end{figure}

The Figures~\ref{fig:12a}, \ref{fig:12b} and \ref{fig:12c} shows the wavelet analysis of the EN, AN and QN areas as explained in the section~\ref{sec:7} %3.7. 

%\bibliography{ms_a}
%\bibliographystyle{aasjournal}

\end{document}